%% file: q-Tasep-v2.tex
\documentclass[11pt,reqno]{amsart}
  \usepackage{amsaddr}

\usepackage{packages}
\usepackage{notations}

\usepackage{tcolorbox}

\input{tikzgraphs}

\makeindex

\begin{document}

\title{Mesoscopic fluctuation theory of particle systems driven by Poisson noise: study of the $q$-TASEP}

\author{Alexandre Krajenbrink}

\address[Alexandre Krajenbrink]{Quantinuum, Partnership House, Carlisle Place, London SW1P 1BX, United Kingdom}
\email{\href{mailto:alexandre.krajenbrink@quantinuum.com}{alexandre.krajenbrink@quantinuum.com}}

\author{Pierre Le Doussal }


\address[Pierre Le Doussal]{Laboratoire de Physique de l’Ecole Normale Sup\'erieure, CNRS, ENS $\&$ Universit\'e PSL, Sorbonne Universit\'e, Universit\'e Paris Cit\'e, 75005 Paris, France}
\email{\href{mailto:pierre.ledoussal@phys.ens.fr}{ledou@lpt.ens.fr}}

\maketitle

\begin{abstract}
We pursue our study of integrable weak noise theories of 
directed polymer and interacting particle stochastic models in the 1D KPZ universality class.
Here we focus on the $q$-TASEP in either continuous or discrete time.
Each particle on $\mathbb{Z}$ jumps independently by $+1$ with a rate (or probability)
depending on the gap to the next particle on its right. We consider initial
conditions (either step or random) which are empty of particles on $\mathbb{Z}^+$,
and focus on the dynamics of the $N$ rightmost particles. 
In the limit $q \to 1$ and at large time (and large gaps) we identify a new intermediate 
"mesoscopic" (i.e. finite $N$) regime which corresponds to weak noise.
In that regime Poisson noise remains important. We obtain the
large deviations of the position of a given particle by two methods.
The first derives asymptotics of $q$-TASEP Fredholm determinant formula. 
The second maps the weak noise limit to a system of semi-discrete or fully discrete,
non linear differential equations. These are obtained as saddle point classical equations 
of a dynamical field theory, and their solutions 
represent the optimal configurations in the large deviation regime. 
We show the classical integrability of these two systems, 
and exhibit their explicit Lax pair. 
In the case of the continuous time $q$-TASEP it provides the first instance of classical integrability arising in a stochastic system, with signatures of the Poisson noise persisting in the weak noise limit. 
For this model, we solve the scattering problem associated to its Lax pair and fully characterize the large deviations associated to the weak noise theory.
Finally, we supplement this work with an Appendix on the first cumulant method
to obtain the large deviations of several lattice polymer models (Strict Weak, Log Gamma, Beta).


\end{abstract}

{\hypersetup{linkcolor=black}
\setcounter{tocdepth}{1}
\makeatletter
\def\l@subsection{\@tocline{2}{0pt}{2.5pc}{2.5pc}{}}
\makeatother
\tableofcontents
}

\section{Introduction and summary of main results}


The weak noise theory of stochastic systems describes situations where the variance of the randomness is small in the natural units, and at the same time the system is conditioned to be in an atypical configuration. Since the 
probability of such an occurrence is exponentially small, this is also called a large deviation regime. The stochastic dynamics bringing the system from its initial state to this atypical configuration becomes dominated by deterministic trajectories \cite{TOUCHETTE20091Review,varadhan1984large,dembo2009large}. These allow to resort to saddle point methods within 
the dynamical field theory (optimal fluctuation theory), and the dynamics can be interpreted as an optimal transport problem, and the trajectory as a geodesics \cite{tsai2023integrability}. The weak noise limit then maps a stochastic field theory
into a classical non linear field theory. One prominent example is 
the so-called macroscopic fluctuation theory (MFT), which provides
a weak noise continuum description of a large class of diffusive interacting systems in one space dimension in terms of stochastic PDE's \cite{bertini2015macroscopic}. The most famous example of such system is the Symmetric Simple Exclusion Process (SSEP) \cite{derrida2009currentMFT,derridareview}.

There has been much recent interest in integrable stochastic models,
in particular within the 1D Kardar-Parisi-Zhang universality class, leading to 
a branch of probability theory called integrable probability \cite{corwin2012kardar,corwin2014macdonald}. 
One of the outcome of integrable probability is to prove exact
formula (usually involving Fredholm determinants)
for some observables.
Provided one identifies adequately some weak noise scaling limits,
it is natural to expect that for each of these integrable stochastic models, 
one can find a classical system which inherits some of their properties.
In particular, it should be integrable in the classical sense, i.e. it should admit a Lax pair representation. 
Identifying these systems for integrable models within the KPZ class
is a program that we started a few years ago, the first example being 
the KPZ equation itself. It was noticed that at short time it can be studied
using optimal fluctuation theory \cite{kolokolov2007optimal,kolokolov2009explicit,meerson2016large},
leading to an
imaginary-time cousin of the non-linear Schrodinger (NLS) equation.
From the integrability of the latter \cite{AblowitzKaup1974,faddeev1987hamiltonian},
we provided a solution of the weak noise theory of the KPZ equation using
inverse scattering methods \cite{UsWNTDroplet2021,UsWNTFlat2021,tsai2023integrability}.
Since then this methodology
has been applied to other models, in particular 
in the realm of the macroscopic fluctuation theory (MFT), where the details of the microscopic noise are encoded in the variance of the Gaussian noise and the diffusion constant of the model. The MFT of the SSEP and of 
the Kipnis-Marchioro-Presutti (KMP) model could be related to the derivative NLS equation \cite{MeersonKMP1,UsWNTCrossover,NaftaliKMP2,SasamotoExactSSEP,grabsch2023exact},
and the MFT of the weakly asymmetric exclusion process (WASEP) to the anisotropic
Landau-Lifschitz equation \cite{krajenbrink2025WASEP}, 
both being integrable non linear PDE's. This allowed
to compute large deviation rate functions as well as 
optimal trajectories for these models. 

This program was later pursued in the realm of discrete and semi-discrete models. The weak noise limit of the O'Connell-Yor (OY) polymer led to an integrable semi-discrete version of the NLS equation \cite{krajenbrink2023weak},
which turns out to be related to the so-called discrete self-trapping
model describing exciton dynamics in molecular crystals \cite{Sklyanin1,enol1991alternate}.
More recently we developed the weak noise theory of 
lattice integrable polymer models such as the Log Gamma, Beta and Strict Weak polymers, and their matrix generalizations \cite{krajenbrinkMatrix}. This led to identify discrete classical
non linear integrable systems which, to our knowledge, are novel discretisations of the non-linear Schrodinger equation.

In the present paper we continue this program to include some models of interacting particles on a lattice.  We study the discrete-time and continuous time $q$-TASEPs \cite{borodin2014duality,borodin2015discrete,corwin2014macdonald,borodin2014macdonald} where particles jump to the right at a rate depending on the gap with the next particle to its right. We consider initial
conditions (either step or random) which are empty of particles on $\mathbb{Z}^+$,
and focus on the dynamics of the $N$ rightmost particles. 
We identify a non-trivial weak noise regime for each model. This regime is obtained in the limit where $q \to 1^-$ while simultaneously 
the gaps between particles become large, scaling as $1/(1-q)$. Hence it describes the motion of the rightmost $N$ particles in a dilute limit. 
Interestingly, for the continuous time $q$-TASEP this weak noise
regime is not described by Gaussian noise, but instead retains some features of the Poisson noise of the original model. This is in contrast with standard examples of macroscopic fluctuation theory (such as for the SSEP) where under diffusive scaling the noise becomes Gaussian. 
The weak noise regime studied here is thus instead a "mesoscopic fluctuation theory" where the fields (i.e. the scaled positions of the particles)
take continuous values, while the labels of the particles (from $1$ to $N$)
remains discrete (and the time remains discrete or continuum). 

Our study of these weak noise regimes, via the saddle point method on the
dynamical action, leads to two novel systems of semi-discrete or fully discrete,
non linear differential equations, see Eqs.~\eqref{eq:sp-continuous-q-tasep} and \eqref{eq:sp-discrete-system-geometric-qtasep} below. We show that these systems are
also classically integrable, by exhibiting their explicit Lax pairs, see Eqs.~\eqref{eq:lax-pair-continuous-q-tasep} and \eqref{eq:lax-pair-discrete-q-tasep} below. 
The scattering theory applied to these systems allows to compute the large deviation rate function which describes the tail of the
distribution of the rescaled positions of the $N$-th rightmost particle. 
This is achieved in details for the step initial condition. 

We also develop a complementary approach based 
on the exact Fredholm determinant formula available
for the $q$-TASEP in discrete and continuous time. 
From the asymptotics of these formula, via the
first cumulant method originally developed in \cite{KrajLedou2018,ProlhacKrajenbrink,krajenbrink2019beyond}, we extract the large deviation rate
functions for step and random initial conditions. For the step initial condition we can check that the result coincides with the scattering approach described above. 

Finally, noting that the structure of the Fredholm determinant formula
are very similar, we complement this work with an Appendix on the first cumulant method
to obtain the large deviations of several lattice polymer models (Strict Weak, Log Gamma, Beta) in their weak noise regimes. Although their associated integrable non-linear systems were already
obtained as a special scalar case in \cite{krajenbrinkMatrix} (with the exception of
the Beta polymer), the present derivation fills a gap as the large deviation rate function have not been previously derived.\\

\paragraph{\textbf{Outline}}

The paper is organized as follows. In Section \ref{sec:continuous}
we study the continuous time $q$-TASEP. We start in 
subsection~\ref{subsec:defcontinuous} by recalling the definition of
the model. In subsection
\ref{subsec:fred-det-qtasep-continuous}
we recall the Fredholm determinant formula for this model
for step and random initial conditions, see Eqs.~\eqref{eq:cont-qtasep-fredholm}-\eqref{eq:first-cum-q-Laplace0}. In subsection~\ref{subsec:weaknoiseFredholm} we define the proper weak noise regime and obtain the asymptotics of the Fredholm determinant formulas in this regime. 
This allows to obtain the rate function of 
of the rescaled positions of the $N$-th rightmost particle.
In subsection~\ref{subsec:dynamicalFT} we derive the
dynamical field theory of the continuous time $q$-TASEP. 
We then study this field theory in the weak noise 
regime and derive saddle point equations in
subsection~\ref{subsec:weaknoiseSP}. In subsection~\ref{subsec:lax-pair-continuous-qtasep} we obtain the Lax pair representation of the saddle point equations, showing
their integrability.

 In Section \ref{sec:discrete}
we study the discrete-time $q$-TASEP by following the same structure. In 
subsection~\ref{subsec:defdiscrete} we recall the definition of
the model and the Fredholm determinant formula (Eq.~\eqref{Kerneldiscrete}) in the case of the step initial condition. In subsection~\ref{subsec:wnt-scaling-discrete} we define the proper weak noise regime and obtain the asymptotics of the Fredholm determinant formulas in this regime yielding the rate function of 
of the rescaled positions of the $N$-th particle.
In subsection~\ref{subsec:dynamicalFT-discrete} we derive the
dynamical field theory of the discrete time $q$-TASEP, derive the weak noise saddle point equations in
subsection~\ref{subsec:weaknoiseSP-discrete} and obtain their Lax pair representation in subsection~\ref{subsec:lax-pair-discrete-qtasep}.

In Section~\ref{sec:scattering-continuous-time-qtasep}, we solve the direct and inverse scattering problem of the continuous time $q$-TASEP with a step initial condition, and derive the conserved quantities of the model in Appendix~\ref{app:conserved}.

We recall in Appendix~\ref{sec:compendium-q-math} the definition of various $q$-deformed combinatorics function. We comment on additional properties of the weak noise theory of the continuous time $q$-TASEP in Appendix~\ref{app:more-qtasep}, notably introducing a generalisation of the Toda chain as a gauge equivalent integrable system to the WNT of the $q$-TASEP and we review the convergence of the $q$-TASEP the O'Connell-Yor polymer through the standpoint of the convergence of the field theory, the weak noise saddle point and the Lax pair. Finally, we detail in Appendix~\ref{app:first-cumulant-method-polymers} the first cumulant method for several lattice polymer models (Strict Weak, Log Gamma and Beta polymers) and summarise the large deviation result for the respective weak noise theory of each polymer model.\\

\paragraph{\textbf{Acknowledgements}}

We thank A.~Aggarwal, A.~Borodin, H.~Desiraju for enlightening discussions and ongoing collaborations.  
PLD acknowledges support from ANR Grant No. ANR-23-CE30-0020-01 EDIPS. We acknowledge partial support from MIT-France MISTI
Global Seed Funds project “Exact Solutions in Field
Theories via Integrable Probability” and the MIT
Mathematics department for hospitality. We thank the Harvard University Center of Mathematical Sciences and Applications, for hospitality and partial support during the 2025 program {\it Classical, Quantum, and Probabilistic Integrable Systems}. 




\section{Continuous time $q$-TASEP}
\label{sec:continuous}

\subsection{Definition of the model}~\\
\label{subsec:defcontinuous}

\begin{figure}[t!]
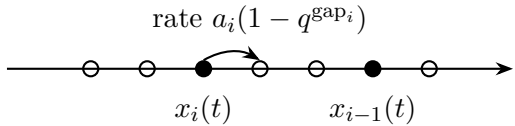

    \centering
    \qTASEPContinuousLattice
    \caption{Representation of the continuous time $q$-TASEP on a one-dimensional lattice. The particles are ordered so that $x_i< x_{i-1}$ for all times and they can only move to the right according to a Poisson clock which rate depends on the gap between consecutive particles.}
    \label{fig:continuous-time-q-tasep}
\end{figure}

The continuous time $q$-TASEP, introduced in Ref.~\cite{borodin2014macdonald}, is a particle system living on a one-dimensional lattice. The positions of the particles, $x_n(t) \in \mathbb{Z}$ are ordered as $x_n(t)<x_{n-1}(t)$ for all $t \geq 0$, with the convention that $x_0(t)=+\infty$. 
Let us define a set of independent Poisson jump process $P_n(t)$,
which increase by $+1$ at random times chosen with independent Poisson clocks of rate $r_n(t)$. 
The evolution of the position of the particles is governed by the semi-discrete Langevin equation with Poisson noise
\begin{equation}
    \rmd \mathsf{x}_n(t)=\rmd P_n(t)
\end{equation}
which amounts to say that the $n$-th particle jumps by $+1$ when its Poisson clock rings. The rate of the Poisson clock is itself time dependent, and is defined as
\begin{equation}
    r_n(t)=a_n (1-q^{\mathsf{x}_{n-1}(t)-\mathsf{x}_n(t)-1})
\end{equation}
for some $q<1$, where $\mathrm{gap}_n=\mathsf{x}_{n-1}-\mathsf{x}_n-1$ is the gap between consecutive particles, and the site-dependent parameters $a_n$ are taken to be positive. The average number of jumps during a time interval $\rmd t$ is given as
\begin{equation}
\big\langle \rmd P_n(t) \big\rangle=r_n(t) \rmd t    
\end{equation}
and the probability that the clock ticks twice during $\rmd t$ is proportional to $(\rmd t)^2$ and thus negligible. We represent in Fig.~\ref{fig:continuous-time-q-tasep} the particle dynamics.
We are interested in the $q$-exponential of the particle position defined as
\begin{equation}
    {\sf z}_n(t)=q^{\mathsf{x}_n(t)+n}
\end{equation}
 Since the position $\mathsf{x}_n(t)$ increases by at most $+1$ during a time interval $\rmd t$, the evolution of the $q$-exponential reads
\begin{equation}
    \rmd {\sf z}_n(t) = (q-1){\sf z}_n(t) \rmd P_n(t)
\end{equation}

As is frequent, we can define the martingale $\rmd M_n(t)=\rmd P_n(t)-\langle \rmd P_n(t) \rangle$ so that the randomness is centered around 0. The dynamics then reads for $n\geq 1$
\begin{equation}
\label{eq:sde-continous-time-q-tasep}
\begin{split}
    \rmd {\sf z}_n    &= a_n(q-1)({\sf z}_n-{\sf z}_{n-1}) \rmd t+(q-1){\sf z}_n \rmd M_n\\
\end{split}
\end{equation}
together with $\mathsf{z}_0(t)=0$. \\

\begin{remark}
The continuous time $q$-TASEP has an interpretation as a model for traffic on a one-lane (discrete) road $\Z$ in which the rate at which cars jump forward is modulated by the distance to the next car. The step initial condition (see \eqref{eq:step-initial-condition-qtasep} below) corresponds to an initially jammed configuration  \cite{borodin2014duality,borodin2015discrete}.
\end{remark}


\begin{remark}[Convergence to the OY polymer] 
    From Ref.~\cite[Eq.~(59)]{borodin2014duality}, the following scaling applied to the $q$-TASEP leads to its convergence to the O'Connell-Yor polymer model
\begin{equation}
\label{eq:scaling-qtasep-to-yo}
    q=e^{-\delta}, \quad t=\delta^{-2}s , \quad \mathsf{x}_n(t) = \delta^{-2}s- (n-1)\delta^{-1}\log \delta^{-1}-\delta^{-1} \mathsf{F}_n(s), \, a_n = e^{-\delta \tilde{a}_n}
\end{equation}
Under the rescaling \eqref{eq:scaling-qtasep-to-yo}, the $q$-deformed partition function becomes
\be 
q^{\mathsf{x}_n+n} ={\mathsf z}_n(t)\to \delta^{1-n} e^{- \frac{s}{\delta} + \mathsf{F}_n(s) }\equiv\delta^{1-n} e^{- \frac{s}{\delta} + \frac{3 }{2}s  }\mathsf{Z}_{n}^{\text{OY}}(s) \, ,
\ee 
where $\mathsf{Z}_{n}^{\text{OY}}(s)$ is the partition sum of the OY polymer,
which satisfies the following stochastic recursion relation, also called
the semi-discrete SHE
\begin{equation} \label{stocheqOY}
    \p_s \mathsf{Z}^{\text{OY}}_n = \mathsf{Z}^{\text{OY}}_{n-1}+(\tilde{a}_n-1)\mathsf{Z}^{\text{OY}}_n+\mathsf{Z}^{\text{OY}}_n \eta_n(s)
\end{equation}
where $\eta_n(s)$ are independent white noises. 
\end{remark}

\subsection{Fredholm determinant formulae} 
\label{subsec:fred-det-qtasep-continuous}

In this Section we recall known Fredholm formulae which will allow
us to obtain the large deviation rate functions subsequently. We first recall that in the $q$-TASEP the first (i.e. rightmost) $N$ particles are independent of the
other particles (to their left).

\subsubsection{Step initial condition} 

We mainly study in this work this particle system from the so-called step initial condition defined as
\begin{equation}
\label{eq:step-initial-condition-qtasep}
    \mathsf{x}_n(t=0)=-n \Longleftrightarrow {\sf z}_n(t=0)=\Theta(n\geq 1), \quad \forall n\geq 1 
\end{equation}
For the step initial condition, the $q$-deformed Laplace transform of ${\sf z}_n$ is described by a Fredholm determinant.  

\begin{theorem}[From Ref.~\cite{borodin2014duality} Theorem 3.12]
\label{the:cont-qtasep-fredholm}
Fix $0<q<1$, $N\geq 1$. Fix $0<\delta<1$, and $\{ a_1,\dots, a_N\}$ such that for all $\ell$, $a_\ell>0$ and $\abs{a_\ell-1}\leq d$, for some constant $d<\frac{1-q^\delta}{1+q^\delta}$. Then for all $t>0$ and $u\in \C/ \R_-$, we have the following
\be 
\label{eq:cont-qtasep-fredholm}
\big\langle \frac{1}{(- u {\sf z}_N(t) ; q)_\infty} \big\rangle = \Det(I+K_u)_{L^2(C_a)} 
\ee  
where $\langle \cdot \rangle$ denotes the expectation value with respect to the stochastic dynamics of the $q$-TASEP, and 
where $K_u$ is the operator of kernel
\be 
\label{eq:fred-det-qmodels}
K_u(v,v') = \int_{\delta + \I \R}\frac{\rmd s}{2 \I \pi} \frac{\pi}{\sin(\pi s)}  u^s \frac{g(v)}{g(q^s v)} \frac{1}{v'-q^s v} 
\ee 
where $C_a$ is a positively oriented contour $|v-1|=d$ 
with measure $\frac{\rmd v}{2 \I \pi }$ and where

    \be 
\label{eq:g-func-continuous-q-TASEP-inhomogeneous}
g(v) =  \prod_{\ell=1}^N\frac{1}{(v/a_\ell, q)_\infty} e^{- t v}  
\ee 
We recall from \eqref{eq:def-q-infinite-pochhammer} the definition of $q$-Pochhammer $(a,q)_{\infty} =\prod_{\ell=0}^{\infty}(1-q^\ell a)$.
\end{theorem}

\subsubsection{Random initial condition}

We consider the following random initial condition (introduced in 
Ref.~\cite{Imamura2019stationaryqtasep}) for the $N$ rightmost particles
\be \label{ic1} 
x_1(0)=-1 \quad , \quad x_{i-1}(0) - x_i(0) - 1 = Y_i \quad , \quad i=2,\dots,N 
\ee 
where $Y_2,\dots,Y_N$ are independent $q$-Poisson random variables with 
parameter $0 \leq \alpha/a_i < 1$. We recall the PDF of the $q$-Poisson distribution
(also called the $q$-geometric distribution)
\begin{equation}
    \mathbb{P}(Y_k=n)= (\alpha/a_k;q)_\infty \frac{(\alpha/a_k)^n}{(q;q)_n}, \quad n \geq 0
\end{equation}
where the finite $q$-Pochhammer reads $(a;q)_{n} = \frac{(a;q)_{\infty}}{(a q^n;q)_{\infty}}$, see \eqref{eq:def-q-finite-pochhammer}. For $q \to 0$ it becomes the geometric distribution of parameter $1-\alpha/a_k$ since 
\begin{equation}
     (\alpha/a_k;q)_\infty \frac{(\alpha/a_k)^n}{(q;q)_n} \to (1-\alpha/a_k)(\alpha/a_k)^n
\end{equation}
while for $q \to 1$
(with $\alpha/a = (1-q)\tilde{\alpha}$) 
it becomes a Poisson distribution since
\begin{equation}
  (\alpha/a_k;q)_\infty \frac{(\alpha/a_k)^n}{(q;q)_n}  \to  e^{-\tilde{\alpha}} \frac{\tilde{\alpha}^n}{n!} \, .
\end{equation}

Note that a more general initial condition 
was considered in \cite{Imamura2019stationaryqtasep} (the initial condition 
\eqref{ic1} corresponds to $\alpha_1=\alpha$
and $\alpha_j=0$ for $j \geq 2$). 

\begin{theorem}[From Ref.~\cite{Imamura2019stationaryqtasep2} Theorem~1]
\label{the:cont-qtasep-fredholm-stationary}
For the random initial condition, the $q$-deformed Laplace transform of ${\sf z}_n$  is described by a Fredholm determinant.    
\be 
\label{eq:first-cum-q-Laplace0}
 \mathbb{E}\left[\Big\langle \frac{1}{(- u {\sf z}_N(t) q^{-Y_1};q)_\infty} \Big\rangle \right]
= \Det(I+K_u)_{L^2(C_a)} 
\ee  
where $u\neq - q^n$, $n \in \Z$,  $C_a$ is a contour including $a_\ell$'s and excluding $\alpha$ 
with measure $\frac{\rmd v}{2 \I \pi }$. There are two expectations values, $\mathbb{E}[\cdot]$ with respect to an additional $q$-Poisson random variable $Y_1$, and $\langle \cdot \rangle$ with respect to the stochastic dynamics of the $q$-TASEP and the initial condition. The kernels reads
\be \label{Kstat} 
K_u(v,v') = \int_{\delta + \I \R}\frac{\rmd s}{2 \I \pi} \frac{\pi}{\sin(\pi s)}  u^s \frac{g(v)}{g(q^s v)} \frac{1}{v'-q^s v} 
\ee 
where 
\be
 g(v) =   \frac{(\alpha/v;q)_\infty}{\prod_{\ell=1}^N (v/a_\ell;q)_\infty}e^{- v t}.
\ee 
This kernel is obtained from \cite[Eq.~(1.10)]{Imamura2019stationaryqtasep2} by a similarity transformation which
leaves invariant the Fredholm determinant.
\end{theorem}

\begin{remark}
Performing a trivial global random shift, i.e. defining $X_n(t)=x_n(t)-Y_1$,
where $Y_1$ is q-Poisson random variable independent from $Y_{n \geq 2}$,
leads to the q-TASEP with a closely related random initial condition for $X_n(t)$, studied in 
\cite{Imamura2019stationaryqtasep} and called half-stationary.
In the case $a_i=1$ for $i \geq 2$ and $a_1 \to \alpha$ one obtains the stationary
initial condition.
\end{remark}

\subsection{Weak noise limit of Fredholm determinant formulae}
\label{subsec:weaknoiseFredholm}

We now study the weak noise limit of these Fredholm determinant formulae.
It will give us large deviation results for the continuous time $q$-TASEP.
We choose the following scaling for the parameters 
\begin{equation}
\label{eq:scaling-wnt-cont-qtasep}
 q=e^{-\varepsilon},  \; t =\frac{\tau}{\varepsilon}, 
  \; \varepsilon \ll 1   , \quad {\sf z}_n(t) \rightarrow {\sf z}_n(\tau) = e^{- \tilde x_n(\tau)}  ,
  \quad {\sf x}_n(t) + n = \frac{\tilde x_n(\tau)}{\varepsilon} 
\end{equation}
where we have introduced the rescaled positions of the particles.
Note that we are also rescaling time. Indeed in this limit we will be interested
in the fluctuations of ${\sf z}_N(t)$ at times $t=T/\varepsilon$, where $T$ is the rescaled observation time. As indicated in \eqref{eq:scaling-wnt-cont-qtasep}, below we supersede the notation ${\sf z}_N(t)$ by ${\sf z}_N(\tau)$.

\subsubsection{Observable of interest: step initial condition} \label{subsubsecobservablestep}
The natural observable for the continuous time $q$-TASEP is the $q$-deformed Laplace transform, which in the case of the step initial condition (on which we focus here) appears in Eq.~\eqref{eq:cont-qtasep-fredholm}. 
It turns out that in the limit of interest here $\varepsilon \to 0$ it can be approximated by an observable which takes a form
suitable to the weak noise large deviations analysis and saddle point methods. 
Using the asymptotics for fixed $x< 1 $
\be \label{asymptoticsPol}
\log (x , q)_{\infty}  \underset{q \to 1}{\to}   -\frac{1}{1-q} {\rm Li}_2(x)+\frac{1}{2}\log(1-x)+\mathcal{O}(1-q)
\ee  
the $q$-deformed Laplace transform is first approximated as\footnote{Here and below,  ${\rm Li}_s$ will denote the polylogarithm with index $s$.}
\be 
\big\langle \frac{1}{(- u {\sf z}_N , q)_\infty} \big\rangle \underset{\varepsilon \ll 1}{\sim}  \big\langle e^{\frac{1}{\varepsilon} {\rm Li}_2(- u {\sf z}_N ) } \big\rangle
\ee  
This gives the asymptotic form of the l.h.s of Eq.~\eqref{eq:cont-qtasep-fredholm}. By evaluating
the asymptotic form of the r.h.s. of Eq.~\eqref{eq:cont-qtasep-fredholm} we show below that
Eq.~\eqref{eq:cont-qtasep-fredholm} takes the following the large deviation
form for $\varepsilon \ll 1$ 
\begin{equation} \label{steplargedev}
    \big\langle  e^{\frac{1}{\varepsilon} {\rm Li}_2(- u {\sf z}_N(T) ) }\big\rangle \underset{\varepsilon \ll 1}{\sim} e^{-\frac{1}{\varepsilon}\Psi_N(u)}
\end{equation}
where the rate function $\Psi_N(u)$ is obtained below in Eq.~\eqref{psiNstep} for the step initial condition. 
    Consequently it is natural to expect that the PDF of $z={\sf z}_N(\tau=T)$ takes the following large deviation form 
    \begin{equation}
        P_N(z)\underset{\varepsilon\ll 1}{\sim} e^{-\frac{1}{\varepsilon}\Phi_N(z)}
    \end{equation}
    and thus we obtain the following Legendre characterisation of the large deviation function
    \begin{equation}
        \Psi_N(u)=\min_{z\in \R_+}[\Phi_N(z)-\mathrm{Li}_2(-uz)]
    \end{equation}
This relation can generally be inverted to obtain parametrically the rate function $\Phi_N(z)$.
Denoting $z(u)$ the argmin, one has 
\be 
\label{eq:continuous-qtasep-legendre-derivative}
\Psi_N'(u)=-\frac{\rmd \, \mathrm{Li}_2(-u z)}{\rmd u}|_{z=z(u)}=\frac{\log(1+u z)}{u}|_{z=z(u)} 
\quad ,\quad z(u) = \frac{e^{u \Psi'_N(u)}-1}{u}  
\ee 
leading to the parametric representation (obtained by varying $u$) 
\begin{equation}
\begin{split}
    \{ \Phi_N(z)=\Psi_N(u)+\mathrm{Li}_2(1-e^{u \Psi'_N(u)}), \quad     z = \frac{e^{u \Psi'_N(u)}-1}{u}  \}
    \end{split}
\end{equation}
Note also that $z \Phi_N'(z) = u \Psi'(u) $.\\

Having obtained the rate function $\Phi_N(z)$ for the 
variable $z$ one easily deduces the one for the scaled 
position $\tilde x= \tilde x_N(T)$ of the $N$-th particle, by
substituting $z=e^{- \tilde x}$. 

\subsubsection{Observable of interest: random initial condition}

We now similarly study the asymptotic form of the formula 
\eqref{eq:first-cum-q-Laplace0} in the limit $\varepsilon \ll 1$
upon the rescaling \eqref{eq:scaling-wnt-cont-qtasep}. 
Using \cite[Appendix~A]{krajenbrink2025WASEP}
the l.h.s. of \eqref{eq:first-cum-q-Laplace0} becomes 
 \be
 \mathbb{E}\left[\Big\langle \frac{1}{(- u {\sf z}_N(T) q^{-Y_1};q)_\infty} \Big\rangle \right] \, 
 \underset{\varepsilon\ll 1}{\sim} \, 
\mathbb{E} \big[ \big\langle  e^{\frac{1}{\varepsilon} {\rm Li}_2(- u \omega {\sf z}_N(T) ) } \big\rangle
\big]
 \ee 
where the expectation value $\mathbb{E}[\cdot]$ is with respect to the
random variable 
$\omega=q^{-Y_1}$ which PDF takes the large deviation form 
\be  
P(\omega)= e^{- \frac{1}{ \varepsilon} F(\omega) } 
 \label{defF} 
 \quad , \quad    F(\omega)=\mathrm{Li}_2\left(1/\omega\right)-\log \omega \log (\alpha/a_1) +\mathrm{Li}_2(\alpha/a_1)-\mathrm{Li}_2(1) \quad , \quad \omega \in [1,+\infty) 
\end{equation}
This PDF has a unique minimum at typical value $\omega_{\rm typ}=\frac{1}{1-\alpha/a_1}$.\\

This gives the asymptotic form of the l.h.s of Eq.~\eqref{eq:first-cum-q-Laplace0}. By evaluating
the asymptotic form of the r.h.s. of ~\eqref{eq:first-cum-q-Laplace0} we show below that
Eq. ~\eqref{eq:first-cum-q-Laplace0} takes the following the large deviation
form for $\varepsilon \ll 1$ 
\begin{equation}
\label{eq:observable-qtasep-continuous-random}
\mathbb{E} \big[ \big\langle  e^{\frac{1}{\varepsilon} {\rm Li}_2(- u \omega {\sf z}_N(T) ) } \big\rangle
\big]
     \underset{\varepsilon \ll 1}{\sim} e^{-\frac{1}{\varepsilon}\Psi_N(u)}
\end{equation}
where the rate function $\Psi_N(u)$ is computed below in Eq.~\eqref{psiN}.\\ 

The observable being more complicated, because of the additional randomness in $\omega$,
the Legendre relation between $\Psi_N(u)$ and $\Phi_N(z)$ now reads
\be 
\Psi_N(u) = \min_{z>0 } [ \Phi_N(z) +  \min_{\omega \in [1,+\infty[} 
[ F(\omega) - {\rm Li}_2(- u \omega z ) ]  ]   \label{PsiMin}
\ee 
The value of $\omega$ which realizes the argmin at fixed $z$ is given by
\begin{equation} \label{omegauz} 
    \omega= \omega_{u,z} = \omega_{u z} \quad , \quad 
    \omega_u = \frac{\alpha/a_1-1+u  +\sqrt{\left(\alpha/a_1 +u -1\right)^2+4 u }}{2 u} 
\end{equation}
Upon differentiation we obtain the following relation at the minimum
\be  \label{uPsi}
\Psi_N'(u)=  \frac{\log (1+u \omega_{uz} z)}{u}   \quad , \quad z \Phi'_N(z) = u \Psi'_N(u) 
\ee 
which allow to reconstruct $\Phi_N(z)$ from the knowledge of $\Psi_N(u)$
(see \cite[Appendix~H]{krajenbrink2025WASEP} for similar manipulations).

\begin{remark}
For $\alpha \to 0$ one recovers the above formula for the step initial condition.
Indeed $\omega \to 1$ in that limit (it becomes deterministic).
\end{remark}

\begin{remark}
The formula \eqref{eq:observable-qtasep-continuous-random} -- \eqref{uPsi} are very similar to those obtain in our recent study of the
WASEP with stationary initial conditions \cite{krajenbrink2025WASEP}.
Both are weak noise limits, respectively of the $q$-TASEP and
of the ASEP. The similarity arises because the $q$-Laplace transforms appear both in the ASEP and in the $q$-TASEP
Fredholm determinant formula. In the weak noise
limit they lead to the same observables \eqref{eq:observable-qtasep-continuous-random}.
The physical quantities analogous to $(z_N(T), \tilde x_N(T),\alpha/a_1)$
in the $q$-TASEP are
 $(z(X,T),2 \nu J(X,T),\alpha)$ in the WASEP, see \cite{krajenbrink2025WASEP}.
Note however that the two models are distinct and the rate functions $\Psi_N(u)$ in the $q$-TASEP and $\Psi(u)$ in the WASEP (given in \cite[Eq.~(5)]{krajenbrink2025WASEP}) are different.
However it is possible to extract the cumulants of $\tilde x_N(T)$ 
by using exactly the same method as detailed
in \cite{krajenbrink2025WASEP} for the cumulants of the current $J(X,T)$.
We will thus not repeat this calculation here.
\end{remark}


\subsubsection{Rate function from the first cumulant method}
\label{subsubsec:first-cum-method-continuous-qtasep}
We now study the asymptotics of $\det(I + K_u)$ in the weak noise limit $\varepsilon \ll 1$.
This is done using the first cumulant method developed in \cite{krajenbrink2025WASEP} for the WASEP, and introduced in \cite{KrajLedou2018,prolhac2009cumulants,krajenbrink2019beyond,krajenbrink2023weak} for the KPZ equation and O'Connell-Yor polymer. Indeed, we note that
the kernels $K_u$ in \eqref{eq:fred-det-qmodels} and \eqref{Kstat}
have exactly the same form as the one for the ASEP that we studied recently in
\cite{krajenbrink2025WASEP}. The only difference is the
function $g(v)$ which depends on the model. We refer to 
\cite[Appendix E]{krajenbrink2025WASEP} for the details of the
derivation (setting $\nu=1/2$, i.e. $\eta=\varepsilon$ there). 
All we need to do here is to obtain the asymptotic form of $g(v)$, i.e. the function $\varphi(v)$
defined by
\be \label{varphi1} 
    \varphi(v)=\lim_{\varepsilon\to 0}\varepsilon\log g(v)|_{q=e^{-\varepsilon}, t=\frac{T}{\varepsilon}} 
\ee 
and then insert it in Eq.~\eqref{psiN}. In the case of random initial condition  for the continuous time $q$-TASEP, the weight $g(v)$ is
\be
 g(v) =   \frac{(\alpha/v;q)_\infty}{\prod_{\ell=1}^N (v/a_\ell;q)_\infty}e^{- v t}.
\ee 
We then obtain, using the asymptotics \eqref{asymptoticsPol} 
\begin{equation}
    \varphi(v) = \lim_{\varepsilon\to 0}\varepsilon\log g(v)|_{q=e^{-\varepsilon}, t=\frac{T}{\varepsilon}} =  - T v +\sum_{\ell=1}^N{ \rm Li}_2(\frac{v}{a_\ell})-{ \rm Li}_2(\frac{\alpha}{v})
\end{equation}
This leads to
\begin{equation}
\begin{split} \label{psiN}
    \Psi_N(u)= \Psi^{(random)}_N(u)&=-\int_{C_a} \frac{\rmd v}{2\I \pi v}\mathrm{Li}_2(-u e^{ v\varphi'(v)})\\
    &=-\int_{C_a} \frac{\rmd v}{2\I \pi v}\mathrm{Li}_2(-u e^{ -v T } \frac{v}{v-\alpha}\prod_{\ell=1}^N \frac{a_\ell}{a_\ell-v})\\
\end{split}
\end{equation}
where we recall that the contour $C_a$ encloses all the $a_i$'s and excludes $\alpha$. This is the result for the random initial condition. The result for
the step initial condition is obtained by setting $\alpha=0$ leading to
\begin{equation}
\begin{split} 
    \Psi_N(u)= \Psi^{(step)}_N(u)    &=-\int_{C_a} \frac{\rmd v}{2\I \pi v}\mathrm{Li}_2(-u e^{ -v T } \prod_{\ell=1}^N \frac{a_\ell}{a_\ell-v})\\
\end{split}
\end{equation}

Note that in the homogeneous case $a_\ell=1$ writing $w=1-v$ one finds
\be 
\label{psiNstep}
\Psi^{(step)}_N(u)  =-\int_{C_0} \frac{\rmd w}{2\I \pi (1-w)}\mathrm{Li}_2(-u w^{-N} e^{ (w-1) T }) 
\ee 
where $C_0$ encloses $0$ but not $1$. Note that 
$\Psi^{(step)}_N(u)|_{T=0}  = - \mathrm{Li}_2(-u)$. 

\begin{remark}
Note the similarity between the rate function of the $q$-TASEP with step initial condition
\begin{equation}
    \p_T \Psi^{(step)}_N(u) = -\int_{C_0} \frac{\rmd w}{2\I \pi }\log(1+u w^{-N} e^{ (w-1) T }) 
\end{equation}
and of the one for the OY polymer with $\{ \tilde{a}_\ell \} =0$ obtained in \cite{krajenbrink2023weak}
\begin{equation}
    \Lambda \p_\Lambda\Psi^{\text{OY}}_N(\Lambda) = \int_{C_0} \frac{\rmd w}{2\I \pi }\log(1+\Lambda w^{-N} e^{ (w-1) \tilde{T} }) 
\end{equation}
\end{remark}

\begin{remark}
[Limit to the weak noise of the OY polymer] Upon performing the rescalings
    \begin{equation} \label{rescalingQ-OY} 
    T= \tilde{T}/\eta, 
    \alpha= e^{-\eta \tilde{\alpha}}, a_i = e^{-\eta \tilde{a}_i},  z_N(T)=\eta^{1-N} e^{- \frac{\tilde{T}}{\eta}+\tilde{T} } Z_{N}^{\text{OY}}(\tilde{T})
\end{equation} 
the weak noise regime of the continuous time $q$-TASEP converges to the weak noise regime of 
the OY polymer, i.e. the partition sum $Z_{n}^{\text{OY}}(\tilde t)$ satisfies 
the semi-discrete SHE with weak noise $\varepsilon/ \eta \ll 1$
\begin{equation}
    \p_{\tilde t} Z^{\text{OY}}_n = Z^{\text{OY}}_{n-1}+(\tilde{a}_n-1)Z^{\text{OY}}_n+ \sqrt{\frac{\varepsilon}{\eta}} 
    Z^{\text{OY}}_n \eta_n(\tilde{t})
\end{equation}
where $\eta_n(\tilde{t})$ are independent white noises. This can be seen on the example of the step initial condition
by considering the limit of 
our large deviation result \eqref{steplargedev} for the $q$-TASEP. 
The left hand side of \eqref{steplargedev} 
becomes
\begin{equation}
    \big\langle  e^{\frac{1}{\varepsilon} {\rm Li}_2(- u {\sf z}_N(T) ) }
    \big\rangle \underset{\eta \ll 1 }{\sim} 
    \big\langle  e^{- \frac{\eta}{\varepsilon} \Lambda  Z_{n}^{\text{OY}}(\tilde{T}) } \big\rangle  , \quad \Lambda = u e^{-T+\tilde{T}}/\eta^{N} 
\end{equation}
while performing the rescaling \eqref{rescalingQ-OY}, together with $v=e^{-\eta \tilde{v}}$,
in Eq. \eqref{psiNstep}
one obtains
\be 
   \Psi^{(step)}_N(u)\underset{\eta \ll 1}{\simeq}  \eta 
   \Psi^{\text{OY}}_N(\Lambda) \quad , \quad 
   \Psi^{\text{OY}}_N(\Lambda) = - 
   \int_{\tilde{C}} \frac{\rmd \tilde v}{2\I \pi}\mathrm{Li}_2\left(- \Lambda e^{ (\tilde v-1) \tilde T } \prod_{\ell=1}^N \frac{1}{\tilde v - \tilde a_\ell}\right)  
\ee 
where $\tilde{C}$ can be chosen as a circle enclosing the $\{\tilde{a}_\ell\}$'s. 
We thus recover the large deviation result for the OY polymer 
\be 
\big\langle e^{- \frac{\eta}{\varepsilon} \Lambda  Z_{n}^{\text{OY}}(\tilde{T}) } \big\rangle 
\underset{\varepsilon/\eta \ll 1}{\sim}  
e^{ - \frac{\eta}{\varepsilon} \Psi^{\text{OY}}_N(\Lambda) }
\ee 
which is identical to the result of
\cite[Eq.~(28)]{krajenbrink2023weak}
in the case $\{\tilde{a}_\ell\}=0$ and $\tilde{T}=1$.


\end{remark}

\subsection{Dynamical field theory}
\label{subsec:dynamicalFT}

We propose to lift the stochastic dynamics \eqref{eq:sde-continous-time-q-tasep} to a dynamical field theory through a Martin-Siggia-Rose approach \cite{andreanov2006field}
by the means of the representation of the generating function of a Poisson noise. Starting from the stochastic semi-discrete equation for the $q$-TASEP \eqref{eq:sde-continous-time-q-tasep}, the MSR approach provides the following path integral representation of the expectation value of any observable of the $q$-TASEP random variables $\{{\sf z} \}$
\begin{equation}
     \big\langle \mathcal{O}(\{{\sf z} \}) \big\rangle =\sum_{\{ z_n \}}  \int \mathcal{D} \tilde{z} \; \big\langle \exp \big( -\sum_{n\geq 1} \int    \tilde{z}_n (\rmd z_n(t) - (q-1)z_n(t) \rmd P_n(t) )\big) \big\rangle \; \mathcal{O}(\{z \})
\end{equation}
where $\langle \cdot \rangle $ denotes the expectation over the Poisson jump process. 
Here $\sum_{\{ z_n \}} $ denotes the sum over all 
configurations $\{z_n(t)\}_{n \in \mathbb{Z}, t \geq 0}$, $\mathcal{D} \tilde{z}$ 
denotes the functional integral measure over the response field 
$\{\tilde z_n(t)\}_{n \in \mathbb{Z}, t \geq 0}$
(which acts as
a Lagrange multiplier to enforce the equation of motion), the
normalization being such that $\langle 1 \rangle = 1$. \\


For a Poisson jump process $P_n(t)$ of unit increments, indexed by an integer $n$ with time-dependent rate $r_n(t)$, the following identity provides its exact generating function \cite{Lefevre2007,andreanov2006field,ken1999levy}
\be 
\label{eq:Poisson-MSR}
\big\langle  e^{\int \rmd P_n(t) F_n(t) } \big\rangle= \big\langle e^{\sum_i F(t_i) } \big\rangle = e^{  \int \rmd t \, r_n(t) (e^{F_n(t)}-1) }
\ee 
where the $\{ t_i \}$ are interpreted as the time when the Poissonian clock ticks. 
This can be understood qualitatively by discretizing time in steps $\Delta t$ as
\begin{equation}
\big\langle  e^{\int \rmd P_n(t) F_n(t) } \big\rangle\sim \prod_t \big\langle  e^{ \rmd P_n(t) F_n(t) } \big\rangle = \prod_t  \sum_{m \geq 0}   \frac{e^{ m F_n(t) } (r_n(t) \Delta t)^m }{m!}e^{-r_n(t) \Delta t}    
\end{equation}
The Poisson noise can be viewed as a
sequence of independent identically distributed pulses arriving in time according to a Poisson counting process. Thus, upon averaging over the Poisson process, the path integral of the $q$-TASEP reads
\begin{equation}
\label{eq:MSR-continuous-q-tasep}
\begin{split}
\big\langle \mathcal{O}(\{{\sf z} \}) \big\rangle    &=\sum_{\{ z_n \}}  \int \mathcal{D} \tilde{z} \exp \left( -\sum_{n\geq 1} \int \rmd t \left[ \tilde{z}_n \p_t  z_n-a_n(z_n-z_{n-1}) \frac{e^{(q-1)z_n\tilde{z}_n}-1}{z_n} \right]\right)\mathcal{O}(\{z \})\\
\end{split}
\end{equation}

The argument of the exponential defines the action as $\exp(-S_0[z,\tilde{z}])$. 

\subsection{Weak noise regime and saddle point equations}
\label{subsec:weaknoiseSP}

\subsubsection{Weak noise action}

Under the scaling \eqref{eq:scaling-wnt-cont-qtasep} together with a rescaling of the response field

\begin{equation}
\label{eq:rescaling-qtasep-wnt}
 q=e^{-\varepsilon},  \; t =\frac{\tau}{\varepsilon}, 
 \; \tilde z_n = \frac{\hat z_n}{\varepsilon} ,
  \; \varepsilon \ll 1   , \quad z_n(t) \rightarrow z_n(\tau)
\end{equation}
Below, we supersede the notation $z_n(t)$ by $z_n(\tau)$.   The path integral measure  becomes at leading order $\iint \mathcal{D}z \mathcal{D}\tilde{z}\exp(-\frac{S[z,\hat{z}]}{\varepsilon})$ with
\be 
\begin{split}
\label{eq:MSR-continuous-q-tasep-rescaled}
S[z,\hat{z}]   &=  \sum_{n\geq 1} \int \rmd \tau \,  \left( \hat z_n \partial_\tau z_n - a_n( z_n-z_{n-1} ) \frac{e^{- z_n \hat z_n } - 1 }{z_n } \right) \\
\end{split}
\ee 
where now $z_n(\tau)$ can be considered as field with values on $\mathbb{R}^+$
and $\mathcal{D}z$ denotes the corresponding path integral measure on 
$\{ z_n(\tau)\}_{n \in \mathbb{Z}, t \geq 0}$. Because of the large
factor $1/\varepsilon$ the path integral is dominated by a saddle point.\\

\begin{remark}
Equation~\eqref{eq:MSR-continuous-q-tasep-rescaled} is thus the dynamical action
of the weak noise theory of the $q$-TASEP. The exponential factor 
$e^{- z_n \hat z_n }$ is a signature that the Poisson noise persists in
the weak noise limit. This feature is new as compared to the weak noise
limits of the KPZ equation \cite{UsWNTDroplet2021} and of the OY polymer \cite{krajenbrink2023weak},
which include Gaussian noise, and can be recovered by expanding this
exponential up to second order.
\end{remark}

To study the saddle point equations, it is more convenient to introduce the following change of variable for the response field $\hat{z}_n \mapsto y_n$
\be
\label{eq:change-var-log-cont-time-qtasep}
1 + y_n z_n = e^{- z_n \hat z_n} \quad \Longleftrightarrow \quad \hat z_n = - \frac{1}{z_n} \log(1+ y_n z_n) =\p_{z_n} \mathrm{Li}_2(-y_n z_n)
\ee 
which transforms the action into its final form
\bea 
\label{eq:action-continuous-q-TASEP}
&& \tilde{S}[z,y]= \sum_{n\geq 1} \int \rmd \tau \,  \left(
- \frac{1}{z_n} \log(1+ y_n z_n) \partial_\tau z_n - a_n( z_n-z_{n-1} ) y_n \right)
\eea 
\begin{remark}
    The Jacobian of the change of variable $\hat{z} \mapsto y$ will be subdominant compared to the rate $1/\varepsilon$ of the path integral and thus we do not include it.
\end{remark}
\begin{remark}
    The time derivative term in the action \eqref{eq:action-continuous-q-TASEP}  is also written as
    \begin{equation}
     - \frac{1}{z_n} \log(1+ y_n z_n) \partial_\tau z_n=\p_{z_n}   (\mathrm{Li}_2(-y_n z_n))\partial_\tau z_n
    \end{equation}
    Under this representation, it is easy to see that the time derivative is symmetric $\tau \to -\tau$ in the two fields
    \begin{equation}
    \label{eq:boundary-term-dilogarithm-time}
        \int \rmd  \tau \p_{z_n}  ( \mathrm{Li}_2(-y_n z_n))\partial_\tau z_n= -\int \rmd  \tau \p_{y_n}   (\mathrm{Li}_2(-y_n z_n))\partial_\tau y_n+ [\mathrm{Li}_2(-y_n z_n)]_{\tau=\tau_i}^{\tau_f}
    \end{equation}
    where the last term is a boundary term.
\end{remark}

\subsubsection{Saddle point equations}

With the choice $\varepsilon \ll 1$, the dynamics will be dominated by the saddle point of the action \eqref{eq:action-continuous-q-TASEP} which yields  the following semi-discrete non-linear system
\be
\label{eq:sp-continuous-q-tasep}
\begin{split}
 \partial_\tau z_n &= a_n(z_{n-1}-z_n) (1 + y_n z_n) \\
  - \partial_\tau y_n &= (a_{n+1}y_{n+1}-a_n y_n) (1 + y_n z_n) 
 \end{split}
\ee
\begin{remark}
    The saddle point equations can additionally be written as
    \begin{equation}
    \begin{split}
       & \p_{z_n}
       \p_{y_n} (\mathrm{Li}_2(-y_n z_n))\p_\tau z_n=a_n(z_n-z_{n-1})\\
       &\p_{z_n}
       \p_{y_n} (\mathrm{Li}_2(-y_n z_n))\p_\tau y_n=a_{n+1}y_{n+1}-a_n y_{n}
    \end{split}
    \end{equation}
\end{remark}

\subsubsection{Boundary conditions} Now that we have obtained the system \eqref{eq:sp-continuous-q-tasep}, it is necessary to understand what the initial and final conditions are. Consider the case of the step initial condition for the $q$-TASEP defined in \eqref{eq:step-initial-condition-qtasep}. It implies the initial condition $z_n(\tau=0)=\Theta(n\geq 1)$ for the system \eqref{eq:sp-continuous-q-tasep}. 

In the weak noise theory, the terminal condition at a final time $\tau=T$ is in one-to-one correspondence with the observable which is studied (see e.g. \cite{krajenbrink2025WASEP,UsWNTDroplet2021,krajenbrink2023weak,kamenev2016short}).
Some observables allow for an explicit solution, depending on the model and the initial condition.
In the present case of the continuous time $q$-TASEP with step initial condition, the natural observable is
\begin{equation} \label{obs} 
\big\langle e^{\frac{1}{\varepsilon} {\rm Li}_2(- u {\sf z}_N(T)  ) } \big\rangle \sim e^{-\frac{1}{\varepsilon}\Psi_N(u)}
\end{equation}
which, as was discussed above, allows to study the large deviations of ${\sf z}_N(\tau=T)$ (note that as in \eqref{eq:rescaling-qtasep-wnt}, $\tau=t/\varepsilon$ denotes scaled time in the weak-noise regime).
The reason why it is the natural observable here, is that, as we show, 
it induces the following terminal condition $y_n(T)=\delta_{nN}u$. From the MSR representation of the $q$-TASEP \eqref{eq:MSR-continuous-q-tasep-rescaled} and the action in the weak noise regime \eqref{eq:action-continuous-q-TASEP}, one has at leading order
\begin{equation}
\big\langle e^{\frac{1}{\varepsilon} {\rm Li}_2(- u {\sf z}_N(T)  ) } \big\rangle \sim \iint \mathcal{D} z \mathcal{D} y \, e^{\frac{1}{\varepsilon} ({\rm Li}_2(- u z_N(T)  ) - \tilde{S}[z,y])}
\end{equation}

The leading order in the exponential (using  \eqref{eq:boundary-term-dilogarithm-time} and noting that we are allowed to discard the boundary term) is

\begin{equation}
\begin{split}
  &{\rm Li}_2(- u z_N(T)  ) - \tilde{S}[z,y]=  {\rm Li}_2(- u z_N  )+\sum_{n \geq 1} \int \rmd \tau \,  \left( \p_{y_n}\mathrm{Li}_2(-y_n z_n)\partial_\tau y_n+ a_n ( z_n-z_{n-1} ) y_n \right)\\
&= \sum_{n\geq 1} \int \rmd \tau \,  \left( \p_{y_n}\mathrm{Li}_2(-y_n z_n)\partial_\tau y_n+ a_n( z_n-z_{n-1} ) y_n +\delta(t-T)\delta_{nN} {\rm Li}_2(- u z_n  )\right)\\
  \end{split}
\end{equation}
We proceed by an integration with respect to the variable $\tau$ in the interval $[T-0^+,T+0^+]$. We use the standard
fact in the MSR representation that the response field vanish beyond the largest observation time from causality, which
implies that $y_n(\tau)=0$ for $\tau>T$. The boundary term of this integration yields
\begin{equation}
   {\rm Li}_2(- y_n z_n  )|_{\tau=T}=\delta_{nN}{\rm Li}_2(- u z_n  )|_{\tau=T}
\end{equation}
which is equivalent to $y_n (T)= \delta_{nN} u$. The weak noise theory of the $q$-TASEP is therefore described by a half-flat problem as we will impose that $z_{+\infty}(\tau)=1$ for all $\tau\geq 0$. In summary, for the observable \eqref{obs} with the step initial condition, the 
system \eqref{eq:sp-continuous-q-tasep} must be solved with the following mixed boundary conditions
\begin{equation}
    \begin{cases}
        z_n(\tau=0)=\Theta(n\geq 1)\\
        y_n (\tau=T)= \delta_{nN} u
    \end{cases}
\end{equation}

\subsection{Lax pair integrability of the weak noise theory of the continuous time $q$-TASEP}
\label{subsec:lax-pair-continuous-qtasep}
In the homogeneous case, i.e., $a_\ell=1$ for all $\ell$, we have found that the system \eqref{eq:sp-continuous-q-tasep} is Lax integrable.

We recall that a semi-discrete non-linear system is said to be Lax integrable \cite{ablowitz2004discrete} if we have the existence of two matrices $L_n(\tau)$ and $U_n(\tau)$ so that the following linear system
obeyed by a vector $v_n(\tau)$
\be 
\label{eq:definition-lax-system-semidiscrete}
v_{n+1}= L_{n} v_{n}  , \quad \p_\tau v_{n}= U_{n} v_{n} \, ,
\ee 
is always compatible. This means that the corresponding relation between the matrices should hold
\be  
\partial_\tau L_n = U_{n+1} L_n - L_n U_n  
\label{eq:comp-semi-discrete} 
\ee 
and should be equivalent to the original semi-discrete non-linear system  \eqref{eq:sp-continuous-q-tasep}. 

\begin{remark}
    The Lax pair is not unique and we can define a gauge transformation $g_{n}$ so that $v_{n}=g_{n}w_{n}$ which modifies the Lax pair as $(L_{n},U_{n})\mapsto(\hat{L}_{n},\hat{U}_{n})$ as
\begin{equation}
\label{eq:gauge-semi-discrete} 
\begin{split}
    \hat{L}_{n}&=g_{n+1}^{-1} L_{n} g_{n}\\
    \hat{U}_{n}&=g_{n}^{-1} U_{n} g_{n}-g_n^{-1}\p_\tau g_n\\
    \end{split}
\end{equation}
while preserving the compatibility equation and the integrable system associated.
\end{remark}

In the present case it turns out that \eqref{eq:sp-continuous-q-tasep} is integrable, $v_n$ is a two-dimensional vector and the matrices $L_n$ and $U_n$ are of size $2\times 2$. The explicit expressions of the
Lax matrices that we obtained are 
\begin{equation}
\label{eq:lax-pair-continuous-q-tasep}
\begin{split}
L_n&
 =\begin{pmatrix}
            1 & 0 \\
            -y_n & 1
        \end{pmatrix}
        \begin{pmatrix}
            \frac{1}{\lambda  \sqrt{y_{n} z_n+1}} & 0 \\
            0 & \lambda  \sqrt{y_{n} z_n+1}
        \end{pmatrix}
             \begin{pmatrix}
            1 & (\lambda^2-1)z_n \\
            0 & 1
        \end{pmatrix}, \\ 
U_{n}&=
    \begin{pmatrix}
 \frac{\lambda ^2-1}{2} +\frac{y_{n} z_{n-1}}{2} & (1-\lambda ^2) z_{n-1} \\
 y_{n} & -\frac{\lambda ^2-1}{2} -\frac{y_{n} z_{n-1}}{2} \\
\end{pmatrix}
\end{split}
\end{equation}
    where $\lambda \in \mathbb{C}$ is the spectral parameter. We have chosen the normalisation of the Lax matrices so that $\Det \, L_n=1$,  $\Tr \, U_n=0$. 

    \begin{remark}
    It remains an open problem to find the Lax pair in the inhomogeneous case.
\end{remark}

    The explicit form of these matrices allows to perform the scattering analysis of the linear system, which we present in Section~\ref{sec:scattering-continuous-time-qtasep} for the step initial condition. The scattering amplitudes then allow to compute the conserved quantities of the non-linear system \eqref{eq:sp-continuous-q-tasep}. From
    them we obtain in Section~\ref{subsec:riemann-hilbert-inverse-scattering} the expression of the large deviation rate function, which we find to be in agreement with the above result 
    \eqref{psiNstep} for the step initial condition obtained from the first 
    cumulant method. 

\section{Discrete time geometric $q$-TASEP}
\label{sec:discrete}
\subsection{Definition of the model}
\label{subsec:defdiscrete}
\begin{figure}[t!]
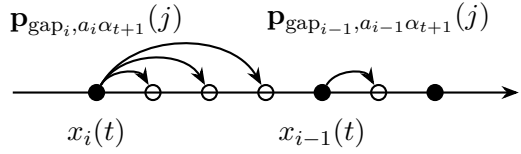

    \centering
    \qTASEPDiscreteLattice
    \caption{Representation of the discrete time $q$-TASEP on a one-dimensional lattice. The particles are ordered so that $x_i< x_{i-1}$ for all times and they can only move to the right according to a random variable which which probability depends on the gap between consecutive particles.}
    \label{fig:discrete-time-qtasep-geometric}
\end{figure}

The discrete time geometric $q$-TASEP,
introduced in Ref.~\cite{borodin2015discrete},
is a discrete version of the continuous time $q$-TASEP. The evolution of the position of the $n$-th particle is given as a discrete random recursion 
\begin{equation}
    \mathsf{x}_{n,t+1}=\mathsf{x}_{n,t}+\mathsf{y}_{n,t}
\end{equation}
where the random variable follows a $q$-deformation of the truncated geometric distribution. The particles evolve using the following parallel rule
\begin{equation}
    \Pr(\mathsf{y}_{n,t}=j | \mathrm{gap}_{n,t})=\mathbf{p}_{\mathrm{gap}_{n,t},a_n \alpha_{t+1}}(j)
\end{equation}
where the gap is given as
\begin{equation}
    \mathrm{gap}_{n,t}=\mathsf{x}_{n-1,t}-\mathsf{x}_{n,t}-1
\end{equation}
and where the updates are independent for each $n$ and $t$ and occur in parallel, see Fig.~\ref{fig:discrete-time-qtasep-geometric}. The probability distribution  reads for $m \geq 0$ and $\alpha \in ]0,1[$

\begin{equation}
\begin{split}
    \mathbf{p}_{m,\alpha}(j)&=\alpha^j (\alpha ;q)_{m-j}\frac{(q;q)_m}{(q;q)_{m-j}(q;q)_j}\mathds{1}\{0 \leq j \leq m\}  \\
    &=\alpha^j (\alpha ;q)_{m-j}\binom{m}{j}_q\mathds{1}\{0 \leq j \leq m\}  \\
\end{split}
\end{equation}
For $m=+\infty$, i.e for the particle in position ${\sf x}_{1,t}$,
the definition can be extended to 
\begin{equation}
\begin{split}
    \mathbf{p}_{+\infty,\alpha}(j)&=\alpha^j \frac{(\alpha ;q)_{\infty}}{(q;q)_j}\mathds{1}\{0 \leq j \}  
\end{split}
\end{equation}
where the $q$-binomial is defined in Eq.~\eqref{eq:def-q-binomial} and more generally we recall the definition of the basic $q$-deformed functions in Appendix~\ref{sec:compendium-q-math}. The parameters $a_n$ and $\alpha_{t+1}$ describe space and time inhomogeneities for the discrete-time $q$-TASEP.
It is assumed through this work that all $a_n$ and $\alpha_t$ are such that $a_n \alpha_t<1$ for all $n$ and $t$. \\

We are interested in the $q$-exponential of the particle position defined as
\begin{equation}
    {\sf z}_{n,t}=q^{\mathsf{x}_{n,t}+n}
\end{equation}

so that
\begin{equation}
    \label{eq:sde-discrete-time-geometric-q-tasep}
    {\sf z}_{n,t+1}={\sf z}_{n,t} q^{y_{n,t}}
\end{equation}

This model has been studied with the step initial condition, identical to the 
one for the continuous time version, and which reads
\begin{equation}
\label{eq:step-initial-condition-qtasep-discrete}
    \mathsf{x}_n(t=0)=-n \Longleftrightarrow {\sf z}_n(t=0)=\Theta(n\geq 1), \quad \forall n\geq 1 
\end{equation}
Let us now recall what is known about this model with this initial condition.

\begin{theorem}[From Ref.~\cite{borodin2015discrete} Theorem~2.4 and Remark~2.5]
\label{the:discrete-qtasep-fredholm}
    Fix $0<q<1$. For the discrete time geometric $q$-TASEP with particle rate parameters $a_n>0$ and time dependent jump parameters $\alpha_t \in ]0,1[$, for all $T\geq 1$, $N \geq 1$ and $u\in \C\setminus \R_-$, we have the following
\be  \label{FDdiscrete}
\big\langle \frac{1}{(- u {\sf z}_{N,T} , q)_\infty} \big\rangle = \Det(I+K_u)_{L^2(C_a)} 
\ee  
with
\be \label{Kerneldiscrete}
K(v,v') = \int_{1/2 + \I \R}\frac{\rmd s}{2 \I \pi} \frac{\pi}{\sin(\pi s)}  u^s \frac{g(v)}{g(q^s v)} \frac{1}{v'-q^s v} 
\ee 
    where $C_a$ is a small positively oriented circle containing the $\{ a_\ell \}$'s while avoiding other poles, the kernel measure is $\frac{\rmd v}{2 \I \pi }$ and where
    \be 
\label{eq:g-func-discrete-q-TASEP-inhomogeneous}
g(v)=\frac{\prod_{s=1}^T(\alpha_s v,q )_\infty}{\prod_{\ell=1}^N(v/a_\ell,q)_\infty }
\ee 
\end{theorem}

\begin{remark}
    There exists a discrete time Bernoulli $q$-TASEP which we do not investigate in this work, see Ref.~\cite{borodin2015discrete}.
\end{remark}

\subsection{Weak noise limit of the Fredholm determinant formula}
\label{subsec:wnt-scaling-discrete}
The weak noise limit of the discrete-time geometric $q$-TASEP is defined simply
by 
\begin{equation}
\label{eq:scaling-wnt-discrete-qtasep}
 q=e^{-\varepsilon} ,  \; \varepsilon \ll 1   
\end{equation}
with no rescaling of space and time, which remain both discrete. We now study the weak noise limit of the Fredholm determinant formula
\eqref{FDdiscrete}, which will give us a large deviation result for the discrete time $q$-TASEP.

\subsubsection{Observable of interest}
Following exactly the same steps as in Section \ref{subsubsecobservablestep}, the observable, i.e. the left hand side of \eqref{FDdiscrete}, becomes in the weak noise limit
\be 
\big\langle \frac{1}{(- u \mathsf{z}_{N,T} , q)_\infty} \big\rangle \underset{\varepsilon \ll 1}{\sim} \big\langle e^{\frac{1}{\varepsilon} {\rm Li}_2(- u \mathsf{z}_{N,T} ) } \big\rangle
\ee  
The Fredholm determinant on the right hand side of \eqref{FDdiscrete} takes
and asymptotic form characterized by the rate function $\Psi_{N,T}(u)$, computed below,
such that at leading order we obtain the large deviation result
\begin{equation}
    \big\langle e^{\frac{1}{\varepsilon} {\rm Li}_2(- u \mathsf{z}_{N,T} ) }\big\rangle \underset{\varepsilon \ll 1}{\sim} e^{-\frac{1}{\varepsilon}\Psi_{N,T}(u)}
\end{equation} 
Similarly, it is natural to expect the following large deviation form for the PDF of $z=\mathsf{z}_{N,T}$
    \begin{equation}
        P_{N,T}(z)\underset{\varepsilon \ll 1}{\sim} e^{-\frac{1}{\varepsilon}\Phi_{N,T}(z)}
    \end{equation}
    and thus we have the following Legendre characterisation of the large deviation function
    \begin{equation}
        \Psi_{N,T}(u)=\min_{z}[\Phi_{N,T}(z)-\mathrm{Li}_2(-uz)]
    \end{equation}
The inversion procedure of this formula is similar to the case of the
continuous time model, see Section \ref{subsubsecobservablestep}.
It is important to note that these formula are valid for any fixed particle label $N$, and any fixed discrete
time $T$, thus it describes the discrete time dynamics of the $N$ rightmost particles.


\subsubsection{Rate function from the first cumulant method}

    


We now compute the rate function $\Psi_{N,T}(u)$ by studying the asymptotics of
the Fredhom determinant in the r.h.s. of \eqref{FDdiscrete} using the first cumulant method. Since the
kernel $K_u$ in \eqref{Kerneldiscrete} has the same form as the one in 
\eqref{eq:fred-det-qmodels} we refer the reader to Section 
\ref{subsubsec:first-cum-method-continuous-qtasep} where this method is discussed. 
The only difference is that for the discrete time $q$-TASEP, from \eqref{eq:g-func-discrete-q-TASEP-inhomogeneous} the weight function is now
\begin{equation}
        g(v)=\frac{\prod_{s=1}^T(\alpha_s v,q )_\infty}{\prod_{\ell=1}^N(v/a_\ell,q)_\infty }
    \end{equation}
In the weak noise limit setting $q=e^{-\varepsilon}$, with $\varepsilon \ll 1$, one defines the function $\varphi(v)$ (analogous to
the one in \eqref{varphi1}) as
\be 
    \varphi(v)=\lim_{\varepsilon\to 0}\varepsilon\log g(v)|_{q=e^{-\varepsilon}} 
\ee 
Using the asymptotics \eqref{asymptoticsPol} we obtain
\begin{equation}
    \varphi(v)= - \sum_{s=1}^T{ \rm Li}_2(\alpha_s v)  +\sum_{\ell=1}^N{ \rm Li}_2(\frac{v}{a_\ell})
\end{equation}
As explained in Section \ref{subsubsec:first-cum-method-continuous-qtasep}, see \eqref{psiN},
the first cumulant method then leads to the following expression for the rate
function of the discrete time $q$-TASEP with step initial conditions
\begin{equation}
\begin{split}
    \Psi_{N,T}(u)&=- \int_{C_a} \frac{\rmd v}{2\I \pi v}\mathrm{Li}_2(-u e^{ v\varphi'(v)})\\
    &=- \int_{C_a} \frac{\rmd v}{2\I \pi v}\mathrm{Li}_2\left(-u \prod_{\ell=1}^N \frac{a_\ell}{a_\ell-v}\prod_{s=1}^T (1-\alpha_s v )\right)\\
\end{split}
\end{equation}
where $C_a$ is small positively oriented circle containing all the $\{ a_\ell\}$'s from Theorem~\ref{the:discrete-qtasep-fredholm}.


\subsection{Dynamical field theory}
\label{subsec:dynamicalFT-discrete}
We now lift the stochastic recursion \eqref{eq:sde-discrete-time-geometric-q-tasep} to a dynamical field theory through a Martin-Siggia-Rose approach using the generating function of the $q$-deformation of the truncated geometric distribution. Starting from the stochastic semi-discrete equation for the $q$-TASEP \eqref{eq:sde-discrete-time-geometric-q-tasep}, the MSR approach provides the following multiple integral representation of the average over an observable of the random variable $\{\mathsf{z}_{n,t} \}$
\begin{equation}
     \big\langle \mathcal{O}(\{{\sf z} \}) \big\rangle =\sum_{\{ z_{n,t} \}}  \int \prod_{n,t} \rmd \tilde{z}_{n,t} \; \big\langle \exp \big( -\sum_{n,t}     \tilde{z}_{n,t}  (z_{n,t+1}-z_{n,t}
q^{\mathsf{y}_{n,t}})\big)\big\rangle \mathcal{O}(\{z \})
\end{equation}
where the overline denotes the expectation over the $q$-deformation of the truncated geometric distribution
for the random variables ${\sf y}_{n,t}$. The expectation value is expressed as
\begin{equation}
\begin{split}
   &\big\langle\exp \left(\tilde{z}_{n,t} z_{n,t}
q^{\mathsf{y}_{n,t}}  \right) \big\rangle= \sum_{j=0}^{m} \exp \left(\tilde{z}_{n,t} z_{n,t}
q^{j}  \right) (a_n \alpha_{t+1})^j (a_n \alpha_{t+1} ;q)_{m-j} \binom{m}{j}_q\\
\end{split}
\end{equation}
with  $m=\mathrm{gap}_{n,t}$.  This sum is equal to a $q$-deformed hypergeometric $_1\phi_1$ function \cite{corwin-qhahn-tasep} which is inconvenient for the asymptotics that we will perform later on. Instead of completely eliminating
${\sf y}_{n,t}$, we keep the noise as a dynamical variable, which we denote $y_{n,t}$,
and write
\begin{equation}
\label{eq:MSR-discrete-q-tasep}
\begin{split}
\big\langle\mathcal{O}(\{{\sf z} \}) \big\rangle    =&\sum_{\{ z_{n,t}, y_{n,t} \}}  \int\prod_{n,t} \rmd \tilde{z}_{n,t} \; \exp \left( -\sum_{n,t}     \tilde{z}_{n,t}  (z_{n,t+1}-z_{n,t}
q^{y_{n,t}})\right)\\
&\times (a_n \alpha_{t+1})^{y_{n,t}} (a_n \alpha_{t+1} ;q)_{\mathrm{gap}_{n,t}-y_{n,t}}\binom{\mathrm{gap}_{n,t}}{y_{n,t}}_q\mathcal{O}(\{z \})
\end{split}
\end{equation}
Exponentiating the integrand allows to define the action as $\exp(-S_0[z,\tilde{z},y])$. In what follows, we also define the $q$-random variable $\omega_{n,t}=q^{\mathsf{y}_{n,t}}$ which has a simple expectation value that is independent of $q$ with our system of variables
  \begin{equation}
    \label{eq:average-discrete-q-tasep-large-dev}
        \langle\omega_{n,t}\rangle=1+  a_n \alpha_{t+1}\left(\frac{z_{n-1,t}}{z_{n,t}}-1\right)
    \end{equation}
We will subsequently supersede the notation $\omega_{n,t}$ to denote the dynamical field equal to $q^{y_{n,t}}$.

 \subsection{Weak noise regime and saddle point equations}
\label{subsec:weaknoiseSP-discrete}
\subsubsection{Weak noise action}

As we now show, under the scaling
\begin{equation}
 q=e^{-\varepsilon} , \; \tilde z_{n,t} = \frac{\hat z_{n,t}}{\varepsilon} , \; q^{y_{n,t}} = \omega_{n,t}
 \; \quad , \quad \varepsilon \ll 1   
\end{equation}
the multiple integral measure in \eqref{eq:MSR-discrete-q-tasep}
takes the form $\exp(-\frac{S[z,\hat{z},\omega]}{\varepsilon})$ at leading order for $\varepsilon \to 0$. To this aim, we first require the following asymptotic\footnote{This asymptotic formula was also used in the context of high energy physics in \cite{faddeev1994quantum}. A more complete series can be found in Ref.~\cite{DilogarithmStackExchange}.} 
for the $q$-Pochhammer as $q \to 1$ to leading order 
\begin{equation}
\label{eq:limit-pochhammer-dilogarithm}
    \log (a;q)_n \underset{\substack{q\to 1,\\ n\to \infty, \\q^n=b }}{\to} -\frac{1}{1-q}\mathrm{Li}_2(a)+\frac{1}{1-q}\mathrm{Li}_2(a b)+\frac{1}{2}\log \left(\frac{1-a}{1-a b} \right)+\mathcal{O}(1-q)
\end{equation}

so that at leading order the various factors in the $q$-deformed truncated geometric distribution read
\begin{equation}
\label{eq:asymptotic-poch-geometric-q-tasep}
\begin{cases}
        \log (q;q)_{\mathrm{gap}_{n,t}} \sim -\frac{1}{1-q}\mathrm{Li}_2(1)+\frac{1}{1-q}\mathrm{Li}_2( \frac{z_{n-1,t}}{z_{n,t}})\\
    \log (q;q)_{y_{n,t}} \sim -\frac{1}{1-q}\mathrm{Li}_2(1)+\frac{1}{1-q}\mathrm{Li}_2( \omega_{n,t})\\
    \log (q;q)_{\mathrm{gap}_{n,t}-y_{n,t}} \sim -\frac{1}{1-q}\mathrm{Li}_2(1)+\frac{1}{1-q}\mathrm{Li}_2( \frac{z_{n-1,t}}{z_{n,t}\omega_{n,t}})\\
    \log  (a_n \alpha_{t+1} ;q)_{\mathrm{gap}_{n,t}-y_{n,t}} \sim -\frac{1}{1-q}\mathrm{Li}_2(a_n \alpha_{t+1} )+\frac{1}{1-q}\mathrm{Li}_2(a_n \alpha_{t+1}  \frac{z_{n-1,t}}{z_{n,t}\omega_{n,t}})
    \end{cases}
\end{equation}
and inserting these asymptotics into Eq. \eqref{eq:MSR-discrete-q-tasep} 
we obtain the weak noise multiple integral representation over three real valued fields
($z,\omega$ being positive) 
\bea 
\big\langle\mathcal{O}(\{{\sf z} \}) \big\rangle  \simeq
 \iiint\prod_{n,t} \rmd \hat{z}_{n,t} \rmd z_{n,t} \rmd \omega_{n,t} 
 \exp(-\frac{S[z,\hat{z},\omega]}{\varepsilon}) \mathcal{O}(\{{\sf z} \})
\eea 
where the weak noise dynamical action reads
\begin{equation}
\label{eq:action-discrete-q-TASEP}
\begin{split}
    S[z,\hat{z},\omega]=&\sum_{n,t} \big[\hat{z}_{n,t} (z_{n,t+1}-z_{n,t} 
\omega_{n,t})+\mathrm{Li}_2(a_n \alpha_{t+1} )-\mathrm{Li}_2(1)+\log (\omega_{n,t})\log(a_n \alpha_{t+1}) \\
&+\mathrm{Li}_2( \omega_{n,t})+\mathrm{Li}_2(  \frac{z_{n-1,t}}{z_{n,t}\omega_{n,t}})-\mathrm{Li}_2( \frac{z_{n-1,t}}{z_{n,t}})-\mathrm{Li}_2(a_n \alpha_{t+1}  \frac{z_{n-1,t}}{z_{n,t}\omega_{n,t}})\big]
\end{split}
\end{equation}

\begin{remark}
    The presence of numerous dilogarithms in the action is reminiscent of terms appearing in identities such as Rogers' Pentagon Identity \cite{Rogers_1907,faddeev1994quantum}.
\end{remark}

\subsubsection{Saddle point equations}

Similarly to the continuous time $q$-TASEP (see Eq.~\eqref{eq:change-var-log-cont-time-qtasep}), we introduce a modified response field $r_{n,t}$ as
\begin{equation}
\label{eq:change-of-variable-response-field-qtasep}
z_{n,t+1} \hat{z}_{n,t}=-\log(1+z_{n,t+1}r_{n,t})
\end{equation}
We now show that the saddle point is described by the following discrete non-linear system
\begin{equation}
\label{eq:sp-discrete-system-geometric-qtasep}
\begin{split}
    z_{n,t+1}-z_{n,t}&=a_n \alpha_{t+1}(z_{n-1,t}-z_{n,t})\frac{1+r_{n,t} z_{n,t}}{1+a_n \alpha_{t+1} r_{n,t} z_{n,t}}\\
r_{n,t-1}-r_{n,t}&=\alpha_{t+1}(a_{n+1} r_{n+1,t}-a_n r_{n,t})\frac{1+r_{n,t} z_{n,t}}{1+a_n \alpha_{t+1}r_{n,t} z_{n,t}}
\end{split}
\end{equation}
To this aim, we study the saddle point with respect to the three fields one by one.
\begin{enumerate}
    \item The saddle point with respect to the response field $\hat{z}_{n,t}$ is
\begin{equation}
\label{eq:q-tasep-discrete-saddle0}
    z_{n,t+1}=z_{n,t}\omega_{n,t}
\end{equation}
\item The saddle point with respect to noise $\omega_{n,t}$ is
\begin{equation}
\label{eq:q-tasep-discrete-saddle1}
    \begin{split}
        \log (a_n \alpha_{t+1}
   ) -\log (1-\omega_{n,t})+\log \left(1-\frac{z_{n-1,t}}{\omega_{n,t} z_{n,t}}\right)-\log \left(1-\frac{a_n \alpha_{t+1}  z_{n-1,t}}{\omega_{n,t} z_{n,t}}\right)=z_{n,t}\hat{z}_{n,t}\omega_{n,t}
    \end{split}
\end{equation}
\end{enumerate}

 The combination of the first two saddle points indicate that the change of variable \eqref{eq:change-of-variable-response-field-qtasep}, i.e., 
 \begin{equation}
z_{n,t+1} \hat{z}_{n,t}=-\log(1+z_{n,t+1}r_{n,t})
\end{equation}
allows to transform the saddle point equation \eqref{eq:q-tasep-discrete-saddle1} into a rational equation after exponentiation. Additionally, from the expectation value of $\omega_{n,t}$ given in Eq.~\eqref{eq:average-discrete-q-tasep-large-dev}, we propose another change of variable for the noise
\begin{equation}
\label{eq:change-of-variable-noise-qtasep}
    \omega_{n,t}=1+ a_n \alpha_{t+1}\left(\frac{z_{n-1,t}}{z_{n,t}}-1\right) (1+\tilde{w}_{n,t})
\end{equation}
so that $\tilde{w}$ represents a random variable with zero average. Upon injecting these change of variables into \eqref{eq:q-tasep-discrete-saddle1}, we find the optimal noise as a function of the field $z_{n,t}$ and the modified response field $r_{n,t}$ as
\begin{equation}
\tilde{w}_{n,t}=\frac{1+r_{n,t} z_{n,t}}{1+a_n \alpha_{t+1} r_{n,t} z_{n,t}}-1
\end{equation}



\begin{enumerate}
  \setcounter{enumi}{2}
  \item Finally, the saddle point with respect to the field $z_{n,t}$ reads

\begin{equation}
\label{eq:q-tasep-discrete-saddle2}
    \begin{split}
        &\log
   \left(1-\frac{z_{n-1,t}}{\omega_{n,t} z_{n,t}}\right)-\log \left(1-\frac{a_n \alpha_{t+1} z_{n-1,t}}{\omega_{n,t} z_{n,t}}\right)+\log \left(1-\frac{a_{n+1} \alpha_{t+1} z_{n,t}}{\omega_{n+1,t}  z_{n+1,t}}\right)-\log \left(1-\frac{z_{n,t}}{\omega_{n+1,t} z_{n+1,t}}\right)\\
        &+z_{n,t} (\hat{z}_{n,t-1}-\omega_{n,t}
   \hat{z}_{n,t})-\log \left(1-\frac{z_{n-1,t}}{z_{n,t}}\right)+\log \left(1-\frac{z_{n,t}}{z_{n+1,t}}\right)=0
    \end{split}
\end{equation}
\end{enumerate}
This equation has the form of a continuity equation. Exponentiating this identity after inserting the changes of variables \eqref{eq:change-of-variable-response-field-qtasep} and \eqref{eq:change-of-variable-noise-qtasep}, we find
\begin{equation}
\label{eq:continuity-eq-q-tasep-discrete1}
\frac{1+r_{n,t-1}z_{n,t}}{1+r_{n,t}z_{n,t+1}}=\frac{1+a_{n+1} \alpha_{t+1}r_{n+1,t}z_{n,t}}{1+a_n \alpha_{t+1}r_{n,t}z_{n-1,t}}
\end{equation}

\begin{figure}[t!]
    \centering
\begin{tikzpicture}[scale=2.6]

\draw[step=2cm,gray,very thin] (0,0) rectangle (1,1);

\node[below] at (0,0) {$z_{n-1,t}$};
\node[below] at (1,0) {$z_{n,t}$};
\node[above] at (1,1) {$z_{n,t+1}$};

\draw[->, dashed, line width=0.6mm, >=latex'] (0,0) --  (1,1);
\draw[->, dashed, line width=0.6mm, >=latex'] (1,0) -- (1,1);

\end{tikzpicture} \hspace{1cm}
\begin{tikzpicture}[scale=2.6]

\draw[step=2cm,gray,very thin] (0,0) rectangle (1,1);

\node[below] at (0,0) {$r_{n,t-1}$};
\node[above] at (1,1) {$r_{n+1,t}$};
\node[above] at (0,1) {$r_{n,t}$};

\draw[->, dotted, line width=0.6mm, >=latex'] (1,1) -- (0,0);
\draw[->, dotted, line width=0.6mm, >=latex'] (0,1) -- (0,0);

\end{tikzpicture}
\caption{Discrete time $q$-TASEP weak noise lattice recurrence dynamics (for $z_{n,t}$ to the left, and $r_{n,t}$ to the right).}
\label{fig:discrete-q-tasep-weak-noise}
\end{figure}
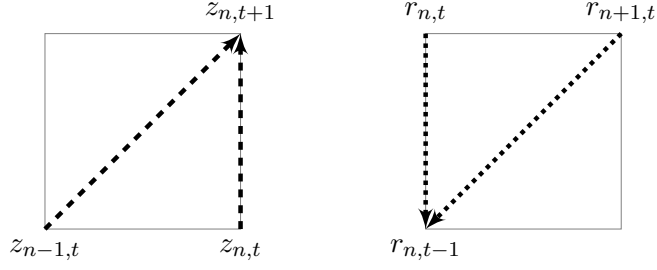

Finally, replacing $z_{n,t+1}$ in \eqref{eq:continuity-eq-q-tasep-discrete1} by the first saddle point equation \eqref{eq:q-tasep-discrete-saddle0} with the change of variable \eqref{eq:change-of-variable-noise-qtasep}, we obtain the system announced in \eqref{eq:sp-discrete-system-geometric-qtasep}. We can further comment on the system  \eqref{eq:sp-discrete-system-geometric-qtasep}:
\begin{itemize}
    \item We rewrite the system as
\begin{equation}
\begin{split}
    z_{n,t+1}-z_{n,t}&=a_n \alpha_{t+1}(z_{n-1,t}-z_{n,t})(1+\tilde{w}_{n,t})\\
r_{n,t-1}-r_{n,t}&=\alpha_{t+1}(a_{n+1} r_{n+1,t}-a_n r_{n,t})(1+\tilde{w}_{n,t})
\end{split}
\end{equation}
and interpret $\tilde{w}_{n,t}$ as a local on-site potential. This recursion can then described as two lattice polymers evolving in opposite time direction as  represented in Fig.~\ref{fig:discrete-q-tasep-weak-noise}.
\item There seems to be a hidden space-time symmetry in this system through the following change of variables
\begin{equation}
    \begin{cases}
        r_{n,t}\to z_{t,n}/a_{n}\\
        z_{n,t} \to r_{t,n}\\
        a_n \to 1/\alpha_{t+1}\\
        \alpha_t \to 1/a_n
    \end{cases}
\end{equation}
and then $n\to t$, $t\to n$ in the indices of the fields. This preserves the on-site potential.
\end{itemize}

\subsubsection{Boundary conditions}

Now that we have obtained the system \eqref{eq:sp-discrete-system-geometric-qtasep}, it is necessary to understand what the initial and final conditions are. Consider the case of the step initial condition, one has the initial condition $z_{n,t=0}=\Theta(n\geq 1)$. With regards to the terminal condition, recall that one studies the large deviations of the following observable
\begin{equation} \label{lddiscrete}
\big\langle e^{\frac{1}{\varepsilon} {\rm Li}_2(- u \mathsf{z}_{N,T}  ) } \big\rangle \sim e^{-\frac{1}{\varepsilon}\Psi_{N,T}(u)}
\end{equation}
We now show that this implies the terminal condition 
\begin{equation}
\label{eq:final-condition-discrete-qtasep-1}
\begin{cases}
    r_{n,T-1}= \delta_{n,N}u, \\
    r_{n,t}=0, \mathrm{for} \, t \geq T
    \end{cases}
\end{equation}
which shows that \eqref{lddiscrete} is indeed a natural observable for 
the system \eqref{eq:sp-discrete-system-geometric-qtasep}.

From the MSR representation of the discrete time $q$-TASEP \eqref{eq:MSR-discrete-q-tasep} and the action in the weak noise regime \eqref{eq:action-discrete-q-TASEP}, one has at leading order at small $\varepsilon$
\begin{equation}
\big\langle e^{\frac{1}{\varepsilon} {\rm Li}_2(- u \mathsf{z}_{N,T}  ) } \big\rangle \underset{\varepsilon \ll 1}{\sim} \iiint\prod_{n,t} \rmd \hat{z}_{n,t} \rmd z_{n,t} \rmd \omega_{n,t} 
 \exp(-\frac{S[z,\hat{z},\omega]-{\rm Li}_2(- u z_{N,T})}{\varepsilon})
 \underset{\varepsilon \ll 1}{\sim} e^{-\frac{1}{\varepsilon}\Psi_{N,T}(u)}
\end{equation}

We can rewrite the source term as
\begin{equation}
  {\rm Li}_2(- u z_{N,T}  )   =\sum_{n\geq 1, t\geq 1} \mathrm{Li}_2(-u \delta_{t,T}\delta_{n,N} z_{n,t})
\end{equation}
which we can include in the action  \eqref{eq:action-discrete-q-TASEP} when taking the saddle point. The consequence of including the source term is that Eq.~\eqref{eq:continuity-eq-q-tasep-discrete1} gets modified to
\begin{equation}
\label{eq:continuity-eq-q-tasep-discrete11}
\frac{1+r_{n,t-1}z_{n,t}}{1+r_{n,t}z_{n,t+1}}=\frac{1+a_{n+1} \alpha_{t+1}r_{n+1,t}z_{n,t}}{1+a_n \alpha_{t+1}r_{n,t}z_{n-1,t}}(1+u \delta_{t,T}\delta_{n,N} z_{n,t})
\end{equation}
while the two other saddle point equations are unchanged. Note that
Jacobians of the various change of variables do not contribute to the
saddle point action at leading order.\\

Assuming that $r_{n,t}$ vanishes for large $t$, Eq.~\eqref{eq:continuity-eq-q-tasep-discrete11} indicates that $T-1$ is the first time where $r_{n,t}$ is non-zero and yields \eqref{eq:final-condition-discrete-qtasep-1}. As was the case for the continuum version, the weak noise theory of the discrete time $q$-TASEP is therefore described by a half-flat scattering problem, as we will impose that $z_{+\infty,t}=1$ for all $t\geq 0$.\\

\begin{figure}[ht!]
\begin{tikzpicture}[scale=0.6]

%
\draw[->, thick,>=latex'] (0, 0) -- (0, 10.3);
\draw[->, thick,>=latex'] (0,0)--( 16.3, 0);



\draw[ultra thick] (0,1) -- (0,9) node[midway, above left] {$t=0$};
\draw[ultra thick] (0,1) -- (7,1) node[midway, below] {$n=1$};

\draw[ultra thick] (0,9) -- (15,9) node[midway, above] {$n=N$};
\draw[ultra thick] (7,1) -- (15,9) node[midway, right, xshift=0.5cm] {$n-N=t-(T-1)$};

\foreach \k in {1,2, ..., 16}
	{\draw[gray, dotted] (\k, 0) -- (\k, 10.1);}
\foreach \k in {1,2, ..., 10}
	{\draw[gray, dotted] (0,\k) -- (16.1, \k);}

\clip (-2, -1) rectangle (18.5, 11.3);

\foreach \k in { 2, 3, ..., 9}
	{\draw[gray] (0,\k) -- (\k+6, \k);}
\foreach \k in { 2, 3, ..., 9}
	{\draw[gray] (0,\k) -- (9-\k, 9);}
\foreach \k in{ 0,1, ..., 7}
	{\draw[gray] (\k, 1) -- (\k+8,9);}

\fill (15,9) circle(3pt);
\fill (0,1) circle(3pt);

\draw(15,9) node[anchor=south]{$(T-1, N)$};
\draw (0,1) node[anchor=east]{$(0,1)$};

\node[above] at (0,10.5) {$n$};
\node[right] at (16.5,0) {$t$};

\end{tikzpicture}
\caption{The black thick lines delimit the intersection of the light cones for the weak noise theory of the discrete time $q$-TASEP. The fact that the upper limit of the light cone is located at $n=N$ indicates that if we consider a $q$-TASEP process with $M>N$ particles and we are interested in the statistics of the $N$-th one, then we can restrict the original process from $M$ to $N$ particle without changing the observable of interest since it does not enter within the light cone.}
\label{fig:discrete-q-tasep-lightcone-wnt}
\end{figure}
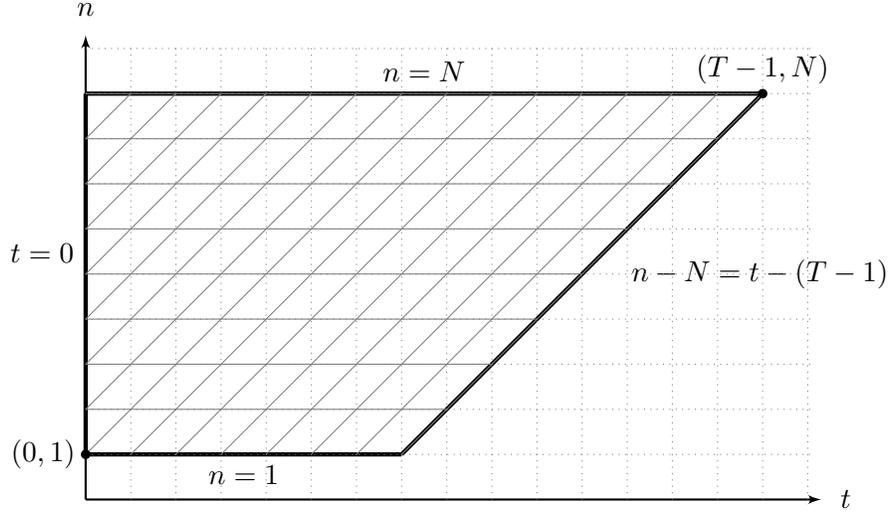

Finally, given the lattice representation of Fig.~\ref{fig:discrete-q-tasep-weak-noise} of the evolution of the system \eqref{eq:sp-discrete-system-geometric-qtasep}, we have represented in Fig.~\ref{fig:discrete-q-tasep-lightcone-wnt} the intersection of the light cones of the forward and backward equations in the case of the step initial condition when the observable is ${\rm Li}_2(- u z_{N,T}  )$.

\subsection{Lax pair integrability of the weak noise theory of the discrete time $q$-TASEP}
\label{subsec:lax-pair-discrete-qtasep}
In the homogeneous case $a_n \alpha_{t+1}=\alpha<1$ for all $n$ and $t$, we have found that the system \eqref{eq:sp-discrete-system-geometric-qtasep} is Lax integrable.

\begin{figure}[ht!]
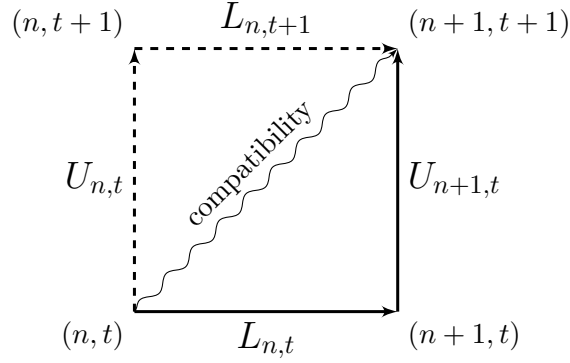

    \centering
    \IntegrabilityCurvature
    \caption{The compatibility of a discrete integrable system is seen from the equivalence of the two paths on the lattice $(n,t)\to (n+1,t) \to (n+1,t+1)$ and $(n,t)\to (n,t+1) \to (n+1,t+1)$. This is the discrete analogous of a zero curvature condition in a continuous space.}
    \label{fig:discrete-Lax-compatibility}
\end{figure}

We recall that a discrete model defined lattice with indices $(n,t)$ is said to be Lax integrable if we have the existence of two matrices $L_{n,t}$ and $U_{n,t}$ so that the following system is compatible
\be 
\label{eq:definition-lax-system-discrete}
v_{n+1,t}= L_{n,t} v_{n,t} , \quad v_{n,t+1}= U_{n,t} v_{n,t} \, .
\ee 
We define the compatibility of the equality between the two ways of reaching $v_{n+1,t+1}$ from $v_{n,t}$ which implies that
\be 
\label{eq:comp-discrete} 
L_{n,t+1} U_{n,t} =U_{n+1,t} L_{n,t}
\ee 
We represent this equality in Figure~\ref{fig:discrete-Lax-compatibility}.

\begin{remark}

For a given discrete integrable model, its associated Lax pair is not unique. Indeed, we can define a gauge transformation $g_{n,t}$ so that $v_{n,t}=g_{n,t}w_{n,t}$ which modifies the Lax pair as $(L_{n,t},U_{n,t})\to(\hat{L}_{n,t},\hat{U}_{n,t})$ as
\begin{equation}
\begin{split}
    \hat{L}_{n,t}&=g_{n+1,t}^{-1} L_{n,t} g_{n,t}\\
    \hat{U}_{n,t}&=g_{n,t+1}^{-1} U_{n,t} g_{n,t}
    \end{split}
\end{equation}
while preserving the compatibility equation and thus the underlying integrable system.
\end{remark}

In the present case it turns out that \eqref{eq:sp-discrete-system-geometric-qtasep}
is integrable, $v_{n,t}$ is a two-dimensional vector and the matrices $L_{n,t}$ and $U_{n,t}$ are of size $2\times 2$. In the simplest homogeneous case $a_n \alpha_{t+1}=\alpha<1$ for all $n$ and $t$, the explicit expressions of the
Lax matrices that we obtained are 
\begin{equation}
\label{eq:lax-pair-discrete-q-tasep}
\begin{split}
L_{n,t}
&=
      \begin{pmatrix}
            1 & 0 \\
            -r_{n,t-1} & 1
        \end{pmatrix}
        \begin{pmatrix}
            \frac{1}{\lambda  \sqrt{1+z_{n,t}r_{n,t-1} }} & 0 \\
            0 & \lambda  \sqrt{1+z_{n}r_{n,t-1} }
        \end{pmatrix}
             \begin{pmatrix}
            1 & (\lambda^2-1)z_{n,t} \\
            0 & 1
        \end{pmatrix}
\end{split}
\end{equation}
and
\begin{equation}
\begin{split}
    &U_{n,t}=\\
&         \begin{pmatrix}
            1 & -\alpha(\lambda^2-1)z_{n-1,t} \\
            0 & 1
        \end{pmatrix}
        \begin{pmatrix}
           \sqrt{1+\alpha z_{n-1,t}r_{n,t}}\sqrt{1+\alpha(\lambda^2-1)}& 0 \\
            0 & \frac{ 1}{\sqrt{1+\alpha z_{n-1,t}r_{n,t}}\sqrt{1+\alpha(\lambda^2-1)}}
        \end{pmatrix}
             \begin{pmatrix}
            1 & 0 \\
            \alpha r_{n,t} & 1
        \end{pmatrix}
    \end{split}
\end{equation}
where $\lambda \in \C$ is the spectral parameter. We have chosen the normalisation of the Lax matrices so that $\Det \, L_{n,t}=1$, $\Det \,  U_{n,t}=1$. The explicit form of these matrices allows to obtain the conserved quantities, and to perform the scattering analysis which we leave for future work.
    


\begin{remark}
    It remains an open problem to extend this Lax pair to the inhomogeneous case.
\end{remark}

\section{Scattering theory of the continuous time $q$-TASEP}
\label{sec:scattering-continuous-time-qtasep}
We focus in this Section on deriving the scattering theory for the weak noise system \eqref{eq:sp-continuous-q-tasep} of the continuous time $q$-TASEP with step initial condition. In the homogeneous case where all space rate parameters $a_n=1$, this systems reads
\be
\label{eq:sp-continuous-q-tasep-homogeneous}
\begin{split}
 \partial_\tau z_n &= (z_{n-1}-z_n) (1 + y_n z_n) \\
  - \partial_\tau y_n &= (y_{n+1}-y_n) (1 + y_n z_n) 
 \end{split}
\ee
We have shown that the system is integrable in the sense of the existence of a Lax pair given in Eq.~\eqref{eq:lax-pair-continuous-q-tasep}. The Lax pair representation of the problem is then composed of two linear systems, one differential in time and one recursive in space, i.e.,
\be 
\label{eq:scattering-linear-continuous-qtasep}
v_{n+1}= L_{n} v_{n} \quad , \quad \p_\tau v_{n}= U_{n} v_{n},
\ee 
We recall from \eqref{eq:lax-pair-continuous-q-tasep} the following set of Lax matrices which are compatible for the continuous time $q$-TASEP
\begin{equation} 
\begin{split}
L_n&=\frac{1}{ \sqrt{1+y_{n} z_n}} 
\begin{pmatrix}
 \frac{1}{\lambda } & \left(\lambda -\frac{1}{\lambda }\right) z_n \\
 -\frac{y_n}{\lambda } & \lambda+\frac{y_n z_n}{\lambda }  \\
\end{pmatrix}, \\ 
U_{n}&=
    \begin{pmatrix}
 \frac{\lambda ^2-1}{2} +\frac{y_{n} z_{n-1}}{2} & (1-\lambda ^2) z_{n-1} \\
 y_{n} & -\frac{\lambda ^2-1}{2} -\frac{y_{n} z_{n-1}}{2} \\
\end{pmatrix} \label{eq:lax-pair-continuous-q-tasep0} 
\end{split}
\end{equation}
we recall that $\lambda \in \C$ is the spectral parameter.\\

The step initial condition for the field $z$ corresponds to
\begin{equation}
\begin{cases}
z_{n}(\tau=0)=\Theta(n\geq 1)\\
y_n (\tau=T)= \delta_{nN} u
\end{cases}
\end{equation}
together with the causality conditions $z_n(\tau<0)=0$, $z_{n \leq 0}(\tau)=0$, $y_n(\tau>T)=0$ and $y_{n \geq N+1}(\tau)=0$. It is therefore a half-flat scattering problem and we have to treat this carefully as we will impose that $z_{+\infty}(\tau)=c$ for all $\tau\geq 0$ (with $c=1$, but we keep $c$ arbitrary for book-keeping).

\subsection{Definition of the scattering amplitudes}

Since 
  \begin{equation}
      U_{+\infty}= \begin{pmatrix}
 \frac{\lambda ^2-1}{2}& (1-\lambda ^2) c \\
 0& -\frac{\lambda ^2-1}{2} \\
\end{pmatrix}
  \end{equation}
we write the two independent solutions of the linear Lax pair problem \eqref{eq:scattering-linear-continuous-qtasep} in the form
 $ {v}_n = e^{\frac{\lambda^2-1}{2}\tau}\phi_n$ with $\phi_n=(\phi_1,\phi_2)^\intercal$ and $ {v}_n = e^{-\frac{\lambda^2-1}{2}\tau}\bar{\phi}_n$, which depend on the three variables $(n,\tau,\lambda)$
 (we often suppress some of this dependence below). We have that 
\begin{equation}
    L_{+\infty}=\begin{pmatrix}
 \frac{1}{\lambda } & \left(\lambda -\frac{1}{\lambda }\right)c  \\
 0 & \lambda \\
\end{pmatrix}, \quad 
    L_{-\infty}=\begin{pmatrix}
 \frac{1}{\lambda } & 0 \\
 0 & \lambda \\
\end{pmatrix}
\end{equation}
As in Ref.~\cite{UsWNTFlat2021,ablowitz2004discrete}, we can choose the solutions 
to behave asymptotically as 
\begin{equation}
    \label{bc01}
    \phi_n  \simeq \lambda^{-n}(1,0)^\intercal, \quad \bar{\phi}_n \simeq \lambda^n 
  (0, -1)^\intercal, \quad n\to -\infty \, .
\end{equation}
At $n\to +\infty$, particular solutions read   $\psi_n  \simeq \lambda^{-n}(1,0)^\intercal $ and $\bar{\psi}_n \simeq \lambda^n 
  (-c, -1)^\intercal$. Each solution should be a particular linear combination of these elementary solutions.
  This allows to define the scattering amplitudes $\{a,\atilde,b,\btilde \}$ as
  \begin{equation}
  \label{eq:def-scattering-amplitude}
    \phi_n  \underset{n\to +\infty}{\simeq }
    \begin{pmatrix}
a(\lambda,\tau)\lambda^{-n}+ c b(\lambda,\tau)\lambda^n  \\
b(\lambda,\tau) \lambda^n
\end{pmatrix}, \quad 
\bar{\phi}_n \underset{n\to +\infty}{\simeq }
    \begin{pmatrix}
\btilde(\lambda,\tau)\lambda^{-n}-c \atilde(\lambda,\tau)\lambda^{n} \\
-\atilde(\lambda,\tau)\lambda^{n}
\end{pmatrix}
\end{equation}


Plugging the solution for $\phi, \bar{\phi}$ at $+\infty$ into the time part of the Lax pair, we find that the time dependence of the scattering amplitudes is as follows

\begin{equation}
\begin{cases}
    a(\lambda,\tau)&=a(\lambda)\\
    \atilde(\lambda,\tau)&=\atilde(\lambda)\\
    b(\lambda,\tau)&=b(\lambda)e^{(1-\lambda^2)\tau}\\
    \btilde(\lambda,\tau)&=\btilde(\lambda)e^{(\lambda^2-1)\tau}
    \end{cases}
\end{equation}
Given the choice of Lax pair, the Wronskian is constant in space and time, and is equal to $W=\phi \wedge \bar{\phi}=-1$, therefore we have the normalisation
\begin{equation}
    a(\lambda)\atilde(\lambda)+b(\lambda)\btilde(\lambda)= 1 
\end{equation}
We expect $a(\lambda)$ to be analytic inside a contour and $\atilde(\lambda)$ outside of that contour as in the case of the weak noise theory of the O'Connell-Yor polymer solved in Ref.~\cite{krajenbrink2023weak}.

\subsection{Solution of the scattering problem for the step initial condition}

We now obtain the solution of the scattering problem associated to the non-linear system, for the 
specific initial and final conditions considered in this paper, that we now recall

\begin{equation}
    z_{n,0}=\Theta(n\geq 1) \quad , \quad  y_n (T)= \delta_{nN} u\label{boundary00} 
\end{equation}
and we recall that $z_{n \leq 0}(t)=0$ and $y_{n \geq N+1}(t)=0$. 
The scattering problem is defined by the recursion, for $n \in \mathbb{Z}$ (the time variable
is implicit) 
\be 
\label{eq:scattcomp}
             \phi_{n+1} = \begin{pmatrix}
  \phi^{(1)}_{n+1} \\
 \phi^{(2)}_{n+1} 
\end{pmatrix}
             =
              \frac{1}{ \sqrt{1+y_{n} z_n}} 
\begin{pmatrix}
 \frac{1}{\lambda } & \left(\lambda -\frac{1}{\lambda }\right) z_n \\
 -\frac{y_n}{\lambda } & \lambda+\frac{y_n z_n}{\lambda }  \\
\end{pmatrix}
  \begin{pmatrix}
  \phi^{(1)}_{n} \\
 \phi^{(2)}_{n} 
\end{pmatrix}
\ee 
together with the same equation for $\bar \phi_n$.

From \eqref{eq:def-scattering-amplitude} the scattering amplitudes are then obtained as
\begin{subequations}
\label{eq:scattering-coefficient-asymptotics}
\bea
   && a(\lambda)=\lim_{n\to \infty} \lambda^n (\phi_n^{(1)}-c\phi_n^{(2)})\\
        &&b(\lambda)e^{(1-\lambda^2)\tau}=\lim_{n\to \infty} \lambda^{-n}\phi_n^{(2)}\\
        &&\atilde(\lambda)=-\lim_{n\to \infty} \lambda^{-n} \bar{\phi}_n^{(2)}\\
        &&\btilde(\lambda)e^{(\lambda^2-1)\tau}=\lim_{n\to \infty}\lambda^n(\bar{\phi}_n^{(1)}-c\bar{\phi}_n^{(2)}) 
    \eea
\end{subequations}


We will now consider the scattering problem successively at the initial time $\tau=0$ and at the final time $\tau=T$.
In each case we consider it first for $\phi$ and then for $\bar \phi$.

\subsubsection{Scattering problem for $ \phi$ at $\tau=T$}

Using that $y_n (\tau=T)= \delta_{nN} u$ one must solve (all implicit time arguments here being set to $\tau=T$)
\begin{equation} \label{ee} 
\begin{split}
& \phi^{(1)}_{n+1}= \frac{1}{ \sqrt{1+ \delta_{nN} u z_n}} \left( \frac{1}{\lambda}\phi^{(1)}_{n} + (\lambda-\frac{1}{\lambda})z_n \phi^{(2)}_{n}\right) \\
& \phi^{(2)}_{n+1}=\frac{1}{ \sqrt{1+ \delta_{nN} u z_n}} \left( - \frac{u}{\lambda }\delta_{n,N}\phi^{(1)}_{n}+ (\lambda +\frac{u}{\lambda }z_n \delta_{n,N} ) \phi^{(2)}_{n} \right)
\end{split}
\end{equation}
Let  us solve first the second equation, using the boundary condition \eqref{bc01} with $\tau=T$. One finds
\begin{equation}
\begin{split}
    \phi_n^{(2)}&=0, \quad  n \leq N\\
    \phi_n^{(2)}&=-\frac{u}{\sqrt{1+u z_N}
    }\lambda^{n-N-2}\phi^{(1)}_{N}, \quad n \geq N+1\\
    \end{split}
\end{equation}
We can now solve the first equation in \eqref{ee} and obtain
\begin{equation}
\begin{split}
    \phi^{(1)}_n &= \lambda^{-n},\quad n\leq N\\
    \phi^{(1)}_{N+1} &= \frac{\lambda^{-N-1}}{\sqrt{1+u z_N}}\\
    \phi^{(1)}_{n} &= \frac{\lambda ^{-n}}{\sqrt{1+u z_N}}-\frac{\left(\lambda ^2-1\right) u }{\sqrt{1+u z_N}}\sum _{k=N+1}^{n-1}  z_k(\tau=T) \lambda ^{2 k-n-2 N-2} ,\quad n\geq N+2\\
\end{split}
\end{equation}
Since the sum in the last term is not convergent when $n\to \infty$, we rewrite it as
\begin{equation}
\begin{split}
 &\sum _{k=N+1}^{n-1}
    z_k(\tau=T) \lambda ^{2k-n-2 N-2} =\sum _{k=N+1}^{n-1}
    (z_k(\tau=T)-c) \lambda ^{2k-n-2 N-2}+c\frac{\lambda ^{-n}-\lambda ^{n-2 N-2}}{1-\lambda ^2}\\
\end{split}
\end{equation}
This allows to rewrite the scattering solution as
\begin{equation}
    \begin{split}
        \phi^{(1)}_{n} &= \frac{\lambda ^{-n}}{\sqrt{1+u z_N}}-\frac{\left(\lambda ^2-1\right) u }{\sqrt{1+u z_N}}\sum _{k=N+1}^{n-1}  (z_k(\tau=T)-c) \lambda ^{2 k-n-2 N-2} +\frac{ u c}{\sqrt{1+u z_N}}(\lambda ^{-n}-\lambda ^{n-2 N-2}),\quad n\geq N+2
    \end{split}
\end{equation}

From the asymptotics for $n \to +\infty$ and \eqref{eq:def-scattering-amplitude}, inserting $\phi^{(1)}_{N}=\lambda^{-N}$, one finds
\begin{subequations}
\bea  
    a(\lambda )&=& \frac{ 1+u c}{\sqrt{1+u z_N}} -\frac{\left(\lambda ^2-1\right) u }{\sqrt{1+u z_N}}\sum _{k=N+1}^{\infty}  (z_k(\tau=T)-c) \lambda ^{2 k-2 N-2} \label{resanew} \\
    b(\lambda) e^{(1-\lambda^2)T}  & = & -\frac{u}{\sqrt{1+u z_N}}\lambda^{-2N-2} \label{resbnew} 
    \eea  
\end{subequations}
where we recall that $z_N=z_N(\tau=T)$.
    \subsubsection{Scattering problem for $ \bar \phi$ at $\tau=T$}

One must solve for $ \bar \phi$ (all implicit time arguments here being set to $\tau=T$)
\begin{equation} \label{ff} 
\begin{split}
& \bar \phi^{(1)}_{n+1}= \frac{1}{ \sqrt{1+ \delta_{nN} u z_n}} \left( \frac{1}{\lambda}\bar \phi^{(1)}_{n} + (\lambda-\frac{1}{\lambda})z_n \bar \phi^{(2)}_{n}\right) \\
& \bar \phi^{(2)}_{n+1}=\frac{1}{ \sqrt{1+ \delta_{nN} u z_n}} \left( - \frac{u}{\lambda }\delta_{n,N}\bar \phi^{(1)}_{n}+ (\lambda +\frac{u}{\lambda }z_n \delta_{n,N} ) \bar \phi^{(2)}_{n} \right)
\end{split}
\end{equation}
Again we start with the second equation, using the asymptotics in \eqref{bc01}. 
We find
\begin{equation}
\begin{split}
    \bar \phi^{(2)}_n &= - \lambda^n   , \quad n\leq N\\
    \bar \phi^{(2)}_n&=   \frac{\lambda^{n-N-1}}{\sqrt{1+uz_N}}\left(  - \frac{u}{\lambda }\bar \phi^{(1)}_{N}- (\lambda +\frac{uz_N}{\lambda } ) \lambda^N  \right), \quad n\geq N+1\\
    \end{split}
\end{equation}
From the $n \to +\infty$ asymptotics and \eqref{eq:def-scattering-amplitude} one finds
\begin{equation} \label{aaaa} 
    \atilde(\lambda)=  -\frac{\lambda^{-N-1}}{\sqrt{1+uz_N}}\left(  - \frac{u}{\lambda }\bar \phi^{(1)}_{N}- (\lambda +\frac{uz_N}{\lambda } ) \lambda^N  \right)
\end{equation}
The solution for $\bar \phi^{(2)}_n$ can then be inserted in the first equation in \eqref{ff}. It reads
\begin{equation}
    \begin{split}
         \bar \phi^{(1)}_{n+1}&= \frac{1}{\lambda}\bar \phi^{(1)}_{n} - (\lambda-\frac{1}{\lambda})z_n  \lambda^n , \quad n\leq N-1\\
         \bar \phi^{(1)}_{N+1}&=\frac{1}{\sqrt{1+u z_N}}\left( \frac{1}{\lambda}\bar \phi^{(1)}_{N} - (\lambda-\frac{1}{\lambda})z_N  \lambda^N  \right) , \\
        \bar \phi^{(1)}_{n+1}&=\frac{1}{\lambda}\bar \phi^{(1)}_{n} + (\lambda-\frac{1}{\lambda})z_n   \frac{\lambda^{n-N-1}}{\sqrt{1+uz_N}}\left(  - \frac{u}{\lambda }\bar \phi^{(1)}_{N}- (\lambda +\frac{uz_N}{\lambda } ) \lambda^N  \right), \quad n\geq N+1\\
    \end{split}
\end{equation}
Taking into account the boundary condition \eqref{bc01}, its solution is found as
\begin{equation}
    \begin{split}
        \bar \phi^{(1)}_{n} &=- (\lambda^2-1) \sum_{k=-\infty}^{n-1} z_k \lambda^{2k-n}  , \quad n \leq N,\\
        \bar \phi^{(1)}_{N+1} &=- \frac{\lambda^2-1}{\sqrt{1+u z_N}} \sum_{k=-\infty}^{N} z_k \lambda^{2k-N-1} ,\\
    \end{split}
\end{equation}
The remaining part of the recursion reads
\begin{equation}
    \begin{split}
         \bar \phi^{(1)}_{n+1}&=\frac{1}{\lambda}\bar \phi^{(1)}_{n} - (\lambda-\frac{1}{\lambda}) \atilde (\lambda)z_n   \lambda^{n}, \quad n\geq N+1\\
    \end{split}
\end{equation}
which we solve as
\begin{equation}
    \begin{split}
        \bar \phi^{(1)}_{n} &=\lambda^{-n+N+1}\bar \phi^{(1)}_{N+1}  - (\lambda^2-1) \atilde (\lambda) \sum_{k=N+1}^{n-1}\lambda^{2k-n}z_k  , \quad n \geq N+2\\
        &= - \frac{\lambda^2-1}{\sqrt{1+u z_N}} \sum_{k=-\infty}^{N} z_k \lambda^{2k-n}  - (\lambda^2-1) \atilde (\lambda) \sum_{k=N+1}^{n-1}\lambda^{2k-n}z_k \\
         &= - \frac{\lambda^2-1}{\sqrt{1+u z_N}} \sum_{k=-\infty}^{N} z_k \lambda^{2k-n}  - (\lambda^2-1) \atilde (\lambda) \sum_{k=N+1}^{n-1}\lambda^{2k-n}(z_k-c)- c \atilde (\lambda) (\lambda ^n-\lambda ^{-n+2 N+2})\\
    \end{split}
\end{equation}
From the asymptotics at $n \to +\infty$ and \eqref{eq:def-scattering-amplitude} one finds
\begin{equation} \label{bbbb} 
    \btilde(\lambda) e^{(\lambda^2-1)T} = - \frac{\lambda^2-1}{\sqrt{1+u z_N}} \sum_{k=-\infty}^{N} z_k \lambda^{2k}  - (\lambda^2-1) \atilde (\lambda) \sum_{k=N+1}^{\infty}\lambda^{2k}(z_k-c)+ c \atilde (\lambda) \lambda ^{2 N+2}
\end{equation}
Inserting the value of $\bar \phi^{(1)}_{N}$ and of $\bar \phi^{(1)}_{N+1}$  in \eqref{aaaa} and 
in \eqref{bbbb} we obtain

\begin{subequations}
\be
       \atilde(\lambda)= \sqrt{1+uz_N}  -\frac{u(\lambda^2-1) }{\sqrt{1+uz_N}} \sum_{k=-\infty}^{N} z_k(\tau=T) \lambda^{2k-2N-2} \label{resatnew}  
       \ee
\be
\btilde(\lambda) e^{(\lambda^2-1)T} =c \atilde (\lambda) \lambda ^{2 N+2} - \frac{\lambda^2-1}{\sqrt{1+u z_N}} \sum_{k=-\infty}^{N} z_k(\tau=T) \lambda^{2k}  - (\lambda^2-1) \atilde (\lambda) \sum_{k=N+1}^{\infty}(z_k(\tau=T)-c)\lambda^{2k}\label{resbtnew}
    \ee
\end{subequations}
where we denote $z_N=z_N(\tau=T)$.

    \begin{remark}
One verifies easily the normalisation $a(\lambda)\atilde(\lambda)+b(\lambda)\btilde(\lambda)=1$ from Eqs.~\eqref{resanew}-\eqref{resbnew}-\eqref{resatnew}-\eqref{resbtnew}.      
    \end{remark}


\subsubsection{Scattering problem for $ \phi$ at $\tau=0$}

One must solve, from \eqref{eq:scattcomp} and the initial condition \eqref{boundary00} 
(all implicit time arguments are now at $\tau=0$) 
\begin{equation}
\begin{split}
& \phi^{(1)}_{n+1}=\frac{1}{\sqrt{1+y_n \Theta(n\geq 1)}}\left(\frac{1}{\lambda}\phi^{(1)}_{n} + (\lambda-\frac{1}{\lambda}) \Theta(n\geq 1)\phi^{(2)}_{n} \right)\\
& \phi^{(2)}_{n+1}=\frac{1}{\sqrt{1+y_n \Theta(n\geq 1)}}\left(-\frac{y_n}{\lambda}\phi^{(1)}_{n}+ (\lambda +\frac{\Theta(n\geq 1)y_n}{\lambda }) \phi^{(2)}_{n}  \right)\label{pair0} 
\end{split}
\end{equation}
Using \eqref{bc01}, the first equation implies that
\begin{equation}
    \begin{split}
        \phi_n^{(1)}&=\lambda^{-n}, \qquad n\leq 1\\
    \end{split}
\end{equation}
The second equation in \eqref{pair0} then leads to 
\bea \label{soluphi01} 
&&     \phi^{(2)}_{n}=-\sum_{k=-\infty}^{n-1}y_k(\tau=0)  \lambda^{n-2-2k} , \quad n\leq 1
\eea
For $n\geq 1$, the recursions can be rewritten as 
\begin{equation}
\begin{split}
& \phi^{(1)}_{n+1}=\frac{1}{\sqrt{1+y_n }}\left(\frac{1}{\lambda}(\phi^{(1)}_{n}-\phi^{(2)}_{n}) + \lambda \phi^{(2)}_{n} \right)\\
& \phi^{(2)}_{n+1}=\frac{1}{\sqrt{1+y_n }}\left(\frac{y_n}{\lambda}(\phi^{(2)}_{n} -\phi^{(1)}_{n})+ \lambda  \phi^{(2)}_{n}  \right)
\end{split}
\end{equation}
We can diagonalise this system by subtracting the two equations 
\begin{equation}
    \begin{split}
        \phi^{(2)}_{n+1}-\phi^{(1)}_{n+1}&=\frac{\sqrt{1+y_n }}{\lambda}(\phi^{(2)}_{n} -\phi^{(1)}_{n}), \quad n \geq 1
    \end{split}
\end{equation}
and thus
\begin{equation}
    \begin{split}
        \phi^{(2)}_{n}-\phi^{(1)}_{n}&=\lambda^{-n+1}(\phi^{(2)}_{1} -\phi^{(1)}_{1})\prod_{k=1}^{n-1}(\sqrt{1+y_k(\tau=0) }), \quad n \geq 2
    \end{split}
\end{equation}
Taking the limit $n \to \infty$, we obtain that
\begin{equation}
\begin{split}
    a(\lambda)&= (1+\sum_{k=-\infty}^{0}y_k(\tau=0)  \lambda^{-2k}) \prod_{k=1}^{\infty}(\sqrt{1+y_k(\tau=0) })\\
    &= (1+\sum_{k=-\infty}^{0}y_k(\tau=0)  \lambda^{-2k}) \sqrt{1+uz_N }\\
\end{split}
\end{equation}
where we went from the first to the second line using the value of the conserved quantity $\tilde{C}_0$ from \eqref{eq:ctilde0-conservedquantity} (in the last equation $z_N=z_N(\tau=T)$).\\

\subsubsection{Scattering problem for $\bar \phi$ at $\tau=0$}

The vector $\bar \phi$ satisfies the same equation as $\phi$ but with different boundary conditions
at infinity, see \eqref{bc01}, \eqref{eq:def-scattering-amplitude} (all implicit time arguments are now at $\tau=0$)
\begin{equation}
\begin{split}
& \bar \phi^{(1)}_{n+1}=\frac{1}{\sqrt{1+y_n \Theta(n\geq 1)}}\left(\frac{1}{\lambda} \bar\phi^{(1)}_{n} + (\lambda-\frac{1}{\lambda}) \Theta(n\geq 1)\bar \phi^{(2)}_{n} \right)\\
&\bar \phi^{(2)}_{n+1}=\frac{1}{\sqrt{1+y_n \Theta(n\geq 1)}}\left(-\frac{y_n}{\lambda} \bar\phi^{(1)}_{n}+ (\lambda +\frac{\Theta(n\geq 1)y_n}{\lambda }) \bar\phi^{(2)}_{n}  \right)\label{eqphib0} 
\end{split}
\end{equation}
Using \eqref{bc01} the first equation gives
\begin{equation}
\begin{split}
    &\bar \phi^{(1)}_{n}=0, \quad  n\leq 1\\
    \end{split}
\end{equation}
Using \eqref{bc01} the second equation in \eqref{eqphib0} can be solved as
 \begin{equation}
 \begin{split}
   \bar \phi^{(2)}_{n}&=-\lambda^n,  \quad n\leq1\\
   \end{split}
 \end{equation}

For $n\geq 1$, the recursions can be rewritten as 
\begin{equation}
\begin{split}
& \bar \phi^{(1)}_{n+1}=\frac{1}{\sqrt{1+y_n }}\left(\frac{1}{\lambda}(\bar \phi^{(1)}_{n}-\bar \phi^{(2)}_{n}) + \lambda \bar \phi^{(2)}_{n} \right)\\
& \bar \phi^{(2)}_{n+1}=\frac{1}{\sqrt{1+y_n }}\left(\frac{y_n}{\lambda}(\bar \phi^{(2)}_{n} -\bar \phi^{(1)}_{n})+ \lambda  \bar\phi^{(2)}_{n}  \right)
\end{split}
\end{equation}
We can diagonalise this system by subtracting the two equations 
\begin{equation}
    \begin{split}
      \bar   \phi^{(2)}_{n+1}-\bar \phi^{(1)}_{n+1}&=\frac{\sqrt{1+y_n }}{\lambda}(\bar \phi^{(2)}_{n} -\bar \phi^{(1)}_{n}), \quad n \geq 1
    \end{split}
\end{equation}
and thus
\begin{equation}
    \begin{split}
        \bar{\phi}^{(2)}_{n}-\bar{\phi}^{(1)}_{n}&=\lambda^{-n+1}(\bar{\phi}^{(2)}_{1} -\bar{\phi}^{(1)}_{1})\prod_{k=1}^{n-1}(\sqrt{1+y_k })\\
        &=-\lambda^{-n+2}\prod_{k=1}^{n-1}(\sqrt{1+y_k(\tau=0) }), \quad n \geq 2\\
    \end{split}
\end{equation}
Taking the $n \to \infty$ limit we find that, using again the value of the conserved quantity $\tilde{C}_0$ from \eqref{eq:ctilde0-conservedquantity}.
\begin{equation}
    \btilde(\lambda)=\lambda^2 \sqrt{1+u z_N}
\end{equation}
where we denote $z_N=z_N(\tau=T)$.

\subsubsection{Summary}

We have therefore obtained that (we denote $z_N=z_N(\tau=T)$ here and below)
\begin{equation}
\begin{split}
    a(\lambda)
    &= (1+\sum_{k=-\infty}^{0}y_k(\tau=0)  \lambda^{-2k}) \sqrt{1+uz_N }\\
    &= \frac{ 1+u c}{\sqrt{1+u z_N}} -\frac{u(\lambda ^2-1)  }{\sqrt{1+u z_N}}\sum_{k=N+1}^{\infty}  (z_k(\tau=T)-c) \lambda ^{2 k-2 N-2} 
\end{split}
\end{equation}

\be
       \atilde(\lambda)= \sqrt{1+uz_N}  -\frac{u(\lambda^2-1) }{\sqrt{1+uz_N}} \sum_{k=-\infty}^{N} z_k(\tau=T) \lambda^{2k-2N-2} \label{resatnew-2}  
       \ee

\bea  
    b(\lambda)   & = & -\frac{u}{\sqrt{1+u z_N}}\lambda^{-2N-2} e^{(\lambda^2-1)T}\label{resbnew-2} 
    \eea  
\be
\btilde(\lambda) =\lambda^2 \sqrt{1+u z_N}
    \ee
together with the normalisation $a(\lambda) \tilde{a}(\lambda) +b(\lambda)\tilde{b}(\lambda)=1$ which is equivalent to

\be \label{G} 
    a(\lambda) \atilde(\lambda)= G(\lambda) 
    = 1 + u \lambda^{-2N} e^{(\lambda^2 -1)T} 
 \ee

Furthermore, we observe that $a(\lambda)$ is expressed as a Taylor series around $\lambda=0$ and is thus analytic inside a contour around $\lambda=0$, while $\tilde{a}(\lambda)$ is expressed as a Laurent series for large $\lambda$ and is thus analytic outside a contour around $\lambda=0$ with 
\be 
\lim_{|\lambda|\to  \infty}\tilde{a}(\lambda)=1/\sqrt{1+u z_N}
\ee 
Note additionally that 
\begin{equation}
\label{eq:value-scattering-1}
    a(\lambda=1)=\frac{1+u}{\sqrt{1+u z_N}}, \quad \tilde{a}(\lambda=1)=\sqrt{1+u z_N}
\end{equation}

\subsection{Solution of the Riemann-Hilbert problem for the scattering amplitudes}
\label{subsec:riemann-hilbert-inverse-scattering}

Having solved the scattering problem we can now obtain the rate function $\Psi_N(u)$.
The consistency check with the result \eqref{psiNstep} of the 
first cumulant method, setting $a_\ell=1$ for all $\ell$,
will be to obtain back the following expression
\begin{equation}
\begin{split} \label{recover} 
    u\Psi_N'(u)  &=\int_{C_1} \frac{\rmd v}{2\I \pi v}\log(1+u   \frac{e^{ -v T }}{(1-v)^N}) = \int_{C'_0} \frac{\rmd w}{2\I \pi }\frac{w}{1-w^2}\log(1+u   e^{ (w^2-1) T } w^{-2N})
\end{split}
\end{equation}
where $C_1$ is a contour around $v=1$ not enclosing $0$. To 
get the second form we changed variables $1-v=w^2$, so that
$w$ can be taken on a contour $C'_0$ around the origin (avoiding the points $w=\pm 1$),
which we choose as a circle centered in $0$ of radius smaller than unity.\\

Since from Eq.\eqref{eq:legendreupsiplog} we have that $u\Psi_N'(u)=\log(1+u z_N)$, 
where we recall that we denote $z_N=z_N(T)$,
in order to obtain the rate function we need to obtain the explicit expression of $\tilde{a}(\lambda)$ and evaluate it at $\lambda=1$ from the knowledge of \eqref{eq:value-scattering-1}.\\

This amount to solving the Riemann-Hilbert problem \eqref{G} 
for $a(\lambda)$ and $\tilde a(\lambda)$. It is very similar to what 
was done in Ref.~\cite[Section X]{krajenbrink2023weak} for the case of the OY polymer - indeed the function $G(\lambda)$ obtained in \eqref{G} is identical to the one in \cite{krajenbrink2023weak}.
Hence we only sketch it here and do not give any details. We only present
the solution in the domain of values for $u$ where no soliton
is present. The extension in presence of solitons can be
performed along the same lines as in \cite[Section X]{krajenbrink2023weak}. \\



Now recall that we expect $a(\lambda)$ to be analytic inside a contour $\mathcal{C}$ (we will
call $\mathcal{D}$ the interior of $\mathcal{C}$) and  $\atilde(\lambda)$ to be analytic outside the contour 
$\mathcal{C}$ in the complementary domain $\mathcal{D}^c$. In practice, the contour $\mathcal{C}$ will be a circle centered around 0 of radius $\mathtt{R}<1$ coinciding with $C'_0$,
in which case $a(\lambda)$ is analytic for for $|\lambda|<\mathtt{R}$
and $\atilde(\lambda)$ is analytic for $|\lambda|>\mathtt{R}$.
From the knowledge of the zeroes of $G(\lambda)$, one can then determine the solution for $a(\lambda)$ and $\atilde(\lambda)$. We assume here that $a(\lambda)$ has no zeroes for $|\lambda|<\mathtt{R}$
and that $\atilde(\lambda)$ has no zeroes for $|\lambda|>\mathtt{R}$, or more generally for $\lambda \in \mathcal{D}$
and $\lambda \in \mathcal{D}^c$ respectively (i.e. absence of solitons).
Taking into account that both functions are even functions of $\lambda$
one has
\begin{itemize}
    \item for $\lambda \in \mathcal{D}$
\begin{equation}
    \begin{split}
        \log {a}(\lambda) =&\oint_{\mathcal{C}}\frac{\rmd w}{2\I \pi}\frac{w}{w^2-\lambda^2}\log {a}(w)\\
        0=&\oint_{\mathcal{C}}\frac{\rmd w}{2\I \pi}\frac{w}{w^2-\lambda^2}\log (\sqrt{1+uz_N}\atilde(w)), \quad \lambda \in \mathcal{D}
    \end{split}
\end{equation}
where in the second equality we have closed the contour at infinity (since $\atilde(\lambda)\sqrt{1+uz_N}$
goes to unity for $|\lambda| \to +\infty$). \\
\item for $\lambda \in \mathcal{D}^c$
\begin{equation}
\label{eq:RH-outside}
    \begin{split}
        \log( \tilde {a}(\lambda)\sqrt{1+uz_N}) =&-\oint_{\mathcal{C}}\frac{\rmd w}{2\I \pi}\frac{w}{w^2-\lambda^2}\log \tilde {a}(w)\\
        0=&\oint_{\mathcal{C}}\frac{\rmd w}{2\I \pi}\frac{w}{w^2-\lambda^2}\log {a}(w), \quad \lambda \in \mathcal{D}^c 
    \end{split}
\end{equation}
\end{itemize}
We have used Cauchy's theorem
    \begin{enumerate}
        \item for a function analytic inside the circle of radius $\mathtt{R}$, we have 
        \begin{equation}
                \label{eq:rule-integration1}\oint_{|w|=\mathtt{R}} \frac{\rmd w}{2\I \pi}  \frac{1}{w -\lambda} f(w) =  f(\lambda)\Theta(\lambda<\mathtt{R})
        \end{equation}
        \item and for a function analytic outside the circle of radius $\mathtt{R}$, we have 
              \begin{equation}
                \label{eq:rule-integration2}
                \oint_{|w|=\mathtt{R}} \frac{\rmd \lambda}{2\I \pi}  \frac{1}{w -\lambda} f(w) = \lim_{w \to +\infty}  f(w)- f(\lambda)\Theta(\, \abs{\lambda}>\mathtt{R})
        \end{equation}
    \end{enumerate}
where $\Theta$ denotes the Heaviside function. \\

Subtracting the equations \eqref{eq:RH-outside} for $\lambda \in \mathcal{D}^c$ and using \eqref{G}, we obtain
\begin{equation}
\label{eq:riemann-hilbert-atilde-explicit}
    \begin{split}
         \log (\atilde(\lambda)\sqrt{1+uz_N}) =&- \oint_{\mathcal{C}}\frac{\rmd w}{2\I \pi}\frac{w}{w^2-\lambda^2}\log (1 + u w^{-2N} e^{(w^2 -1)T})  \quad , \quad \lambda \in \mathcal{D}^c 
         \\
    \end{split}
\end{equation}
Since $\lambda=1 \in \mathcal{D}^c $, we finally have that
\begin{equation}
\label{eq:riemann-hilbert-atilde-explicit-2}
    \log (\atilde(\lambda=1)\sqrt{1+uz_N})= \log(1+uz_N)= \oint_{\mathcal{C}}\frac{\rmd w}{2\I \pi}\frac{w}{1-w^2}\log (1 + u w^{-2N} e^{(w^2 -1)T}) )
\end{equation}
which recovers \eqref{recover}.

\section{Conclusion and Outlook}

In this work we have constructed and studied the weak noise theory of the $q$-TASEP, a paradigmatic model of particles driven by Poisson noise in the continuous time case. As the weak noise limit keeps a strong signature of the Poisson noise and of the discreteness of the particles, this limit should be interpreted as a mesoscopic rather than macroscopic
fluctuation theory. 
We have approached the weak noise regime through the lens of two methods: (i) an exact dynamical field theory which leads to integrable saddle point equations
(ii) the first cumulant method, which makes use of the exact Fredholm determinant representation for some observable of the microscopic model. 

Our main contribution, both in the continuous and discrete time setting, lie in the identification of the mesoscopic weak-noise scaling which describes the edge of the gas of particles, the derivation of the saddle-point equations, together with their Lax pair representation, the solution of the scattering problem for the continuous time model and the calculation of the large deviation rate functions in both cases.  


In this paper we focused on specific initial conditions (step and random). One could further
study some universal properties of these systems - independent of the initial condition - such as solitons. These solitons have been shown to play an important role
in the universality of the large deviation tails for the weak noise theory of the KPZ equation, see e.g. \cite{meerson-landau,meerson2016large,kamenev2016short,UsWNTDroplet2021}.
Their study for the $q$-TASEP is still open.

\begin{figure}[h!]
    {\centering
\hspace{-1.5cm}
\begin{tikzpicture}[node distance=0.8cm, auto, scale=0.88]

\draw[dashed] (-9,2.5) -- (7,2.5);
\draw[dashed] (-9,-6.3) -- (7,-6.3);

\node[rotate=55, align=center] at (-9,5) {Discrete Time \\ Discrete Space};
\node[rotate=55, align=center] at (-9,-1.5) {Continuous Time \\ Discrete Space};
\node[rotate=55, align=center] at (-9,-8) {Continuous Time \\ Continuous Space};


\node[rectangle, draw=black, fill=white, rounded corners, align=center, shading=axis, left color=white, right color=white] (qHahnPushTASEP) at (-7.5,7.5) {$q$-Hahn \\ PushTASEP \\ discrete time};

\node[rectangle, draw=black, fill=white, rounded corners,  below =4cm of qHahnPushTASEP, align=center, shading=axis, left color=white, right color=white] (qHahnPushTASEPcontinuous) {$q$-Hahn \\ PushTASEP \\ continuous time};

\node[rectangle, draw=black, fill=white, rounded corners,  right =1.5cm of qHahnPushTASEP, align=center, shading=axis, left color=white, right color=white] (qPushASEP) {$q$-PushASEP \\ discrete time};

\node[rectangle, draw=black, fill=white, rounded corners,  below =4.5cm of qPushASEP, align=center, shading=axis, left color=white, right color=white] (qPushASEPcontinuous) {$q$-PushASEP \\ continuous time};

\node[rectangle, draw=black, fill=white, rounded corners, below right=0.5cm and -0.5cm of qHahnPushTASEP, align=center, shading=axis, left color=white, right color=white] (qPushTASEP) {$q$-PushTASEP  \\ discrete time};

\node[rectangle, draw=black, fill=white, rounded corners,  below =4.8cm of qPushTASEP, align=center, shading=axis, left color=white, right color=white] (qPushTASEPcontinuous) {$q$-PushTASEP \\ continuous time};

\node[rectangle, draw=black, fill=white, rounded corners,  right =4.5cm of qPushASEP, align=center, shading=axis, left color=white, right color=white] (qHahnTASEP) {$q$-Hahn \\ TASEP \\ discrete time};

\node[rectangle, draw=black, fill=white, rounded corners,  below =0.5cm of qHahnTASEP, align=center, shading=axis, left color=gray!50, right color=white] (qTASEP) {$q$-TASEP \\ discrete time};

\node[rectangle, draw=black, fill=white, rounded corners,  below left =0.5cm and 0.5cm of qHahnTASEP, align=center, shading=axis, left color=white, right color=white] (Beta) {Beta \\ polymer};

\node[rectangle, double distance=0.5mm, 
    dashed, 
    dash pattern=on 3pt off 2pt, 
    draw=black, fill=white, rounded corners,  below =0.5cm of Beta, align=center, shading=axis, left color=white, right color=white] (StrictWeak) {Strict weak \\ polymer};

\node[rectangle, draw=black, fill=white, rounded corners,  left =0.5cm of Beta, align=center] (InverseBeta) {Inverse Beta \\ polymer};

\node[rectangle, double distance=0.5mm, 
    dashed, 
    dash pattern=on 3pt off 2pt, 
draw=black, fill=white, rounded corners,  below =0.5cm of InverseBeta, align=center, shading=axis, left color=white, right color=white] (LogGamma) {Log Gamma \\ polymer};

\node[rectangle, draw=black, fill=white, rounded corners, below left =4.2cm and -1.5cm  of qHahnTASEP, align=center, shading=axis, left color=white, right color=white] (qHahnASEPContinuous) {$q$-Hahn ASEP \\ continuous time};

\node[rectangle, draw=black, fill=white, rounded corners, below right=0.3cm and 0.8cm of qHahnASEPContinuous, align=center, shading=axis, left color=white, right color=white] (qHahnTASEPContinuous) {$q$-Hahn TASEP \\ continuous time};

\node[rectangle, draw=black, fill=white, rounded corners, below=0.3cm of qHahnTASEPContinuous, align=center, shading=axis, left color=gray!50, right color=white] (qTASEPContinuous) {$q$-TASEP \\ continuous time};


\node[rectangle, double distance=0.5mm, 
    dashed, 
    dash pattern=on 3pt off 2pt, 
draw=black, fill=white, rounded corners, below=5.5cm of LogGamma, align=center] (OConnellYor) {O'Connell-Yor \\ polymer};

\node[rectangle, draw=black, fill=white, rounded corners, left=2.4cm of qTASEPContinuous, align=center, shading=axis, left color=gray!50, right color=white] (qToda) {$q$-Toda \\ chain};

\node[rectangle, double distance=0.5mm, 
    dashed, 
    dash pattern=on 3pt off 2pt, 
draw=black, fill=white, rounded corners, below=0.9cm of qToda, align=center, shading=axis, left color=white, right color=white] (OYToda) {OY-Toda \\ chain};

\node[rectangle, draw=black, fill=white, rounded corners, below=1cm of qPushTASEPcontinuous, align=center, shading=axis, left color=white, right color=white] (ASEP) {ASEP};

\node[rectangle, draw=black, fill=white, rounded corners, below=0.7cm of ASEP, align=center, shading=axis, left color=white, right color=white] (SSEP) {SSEP};

\node[rectangle, double distance=0.5mm, 
    dashed, 
    dash pattern=on 3pt off 2pt, 
    draw=black, fill=white, rounded corners, below =0.4cm of OYToda, align=center] (Toda) {classical  \\ Toda chain};


\node[rectangle,double, 
    double distance=0.5mm, 
    dashed, 
    dash pattern=on 3pt off 2pt, 
    draw=black, fill=white, rounded corners, shading=axis, left color=white, right color=white, below = 4cm of OConnellYor, align=center] (KPZ)  {KPZ equation \\ (NLS)};

\node[rectangle, double distance=0.5mm, 
    dashed, 
    dash pattern=on 3pt off 2pt, 
draw=black, fill=white, rounded corners,  shading=axis, left color=white, right color=white,  below right=1.7cm and -0.4cm of SSEP, align=center] (MFTSSEP) {MFT of  the SSEP \\ (DNLS)};

\node[rectangle, draw=black, fill=white, rounded corners,  left =0.5cm of MFTSSEP, shading=axis, left color=white, right color=white] (qPNG) {$q$-PNG};

\node[rectangle,double, 
    double distance=0.5mm, 
    dashed, 
    dash pattern=on 3pt off 2pt, 
    draw=black, fill=white, rounded corners, shading=axis, left color=white, right color=white,right=0.8cm of MFTSSEP](WASEP) {WASEP};


\node[rectangle, double distance=0.5mm, 
    dashed, 
    dash pattern=on 3pt off 2pt, 
draw=black, fill=white, rounded corners, shading=axis, left color=white, right color=white, right =1cm of KPZ, align=center] (LandauLifschitz) {Landau-Lifschitz \\ model};


\begin{pgfonlayer}{main}
\draw[-latex] (qHahnPushTASEP.south) -- (qPushTASEP.north);
\draw[-latex] (qHahnPushTASEP) -- (qHahnPushTASEPcontinuous.north);
\draw[-latex] (qPushASEP.south) -- (qPushTASEP.north);
\draw[-latex] (qPushTASEP) -- (qPushTASEPcontinuous);
\draw[-latex] (qPushTASEP.south) -- (LogGamma.north);
\draw[-latex] (qPushTASEPcontinuous) -- (OConnellYor.north);
\draw[-latex] (qPushASEP.south) -- (qPushASEPcontinuous);
\draw[-latex] (qPushASEPcontinuous.south) -- (qPushTASEPcontinuous.north);
\draw[-latex] (qHahnPushTASEPcontinuous.south) -- (qPushTASEPcontinuous.north);
\draw[-latex] (qPushASEP.south) -- (qTASEP.north);
\draw[-latex] (qHahnTASEP.south) -- (Beta.north);
\draw[-latex] (qTASEP.south) -- (StrictWeak.north);
\draw[-latex] (Beta) -- (StrictWeak);
\draw[-latex] (qHahnASEPContinuous.south) -- (qHahnTASEPContinuous.north);
\draw[-latex] (qHahnTASEPContinuous) -- (qTASEPContinuous);
\draw[latex-latex] (qTASEPContinuous.west) -- (qToda.east);
\draw[-latex] (qToda.south) -- (OYToda.north);
\draw[-latex] (OYToda.south) -- (Toda.north);
\draw[-latex]   (InverseBeta) -- (LogGamma);
\draw[-latex]   (InverseBeta.south) -- (StrictWeak.north);
\draw[-latex] (LogGamma) -- (OConnellYor);
\draw[-latex] (StrictWeak.south) -- (OConnellYor.north);
\draw[-latex] (qHahnTASEP) -- (qTASEP);
\draw[-latex] (qHahnTASEP.east) -- (qHahnTASEPContinuous.north);
\draw[-latex] (qTASEP.south) -- (qTASEPContinuous.west);
\draw[-latex] (qTASEPContinuous.south) -- (OConnellYor.north);
\draw[latex-latex] (OConnellYor.east) -- (OYToda.west);
\draw[-latex] (MFTSSEP.south) -- (KPZ.north);
\draw[-latex] (qPNG.south) -- (KPZ.north);
\draw[-latex] (OConnellYor.south) -- (KPZ.north);
\draw[latex-latex] (LandauLifschitz.west) -- (KPZ.east);
\draw[-latex] (ASEP.south) -- (SSEP.north);
\draw[-latex] (ASEP.south) -- (WASEP.north);
\draw[-latex] (WASEP.south) -- (KPZ.north);
\draw[-latex] (SSEP.south) -- (MFTSSEP.north);
 \end{pgfonlayer}

\begin{pgfonlayer}{background}
\draw[-latex] (qPushTASEP) -- (qPNG);
\draw[-latex] (qHahnTASEPContinuous.south) -- (OConnellYor.north);
\end{pgfonlayer}

\end{tikzpicture}%

}
    \caption{Some models in the KPZ universality class. The double-dashed cells indicate the models whose weak noise theory has been already investigated and the shaded cells indicate the models studied in this work. The descending arrows indicate the convergence of models (we have omitted on this figure the models related by a $q\to 0$ limit). 
    Models connected by a horizontal arrow are related through a change of variable (or equivalently a gauge transformation). Note additionally that the MFT of the SSEP is gauge equivalent to the weak noise theory of the KPZ equation (through a non-local transformation discovered by Wadati and Sogo in \cite{Wadati_1983}) and also converges to the WNT of the KPZ equation (through a limit involving a boost transformation), see Ref.~\cite{UsWNTCrossover} for more details. This list is not extensive and notably omits colored and vertex models.} 
    \label{fig:phylogenetic-tree}
\end{figure}
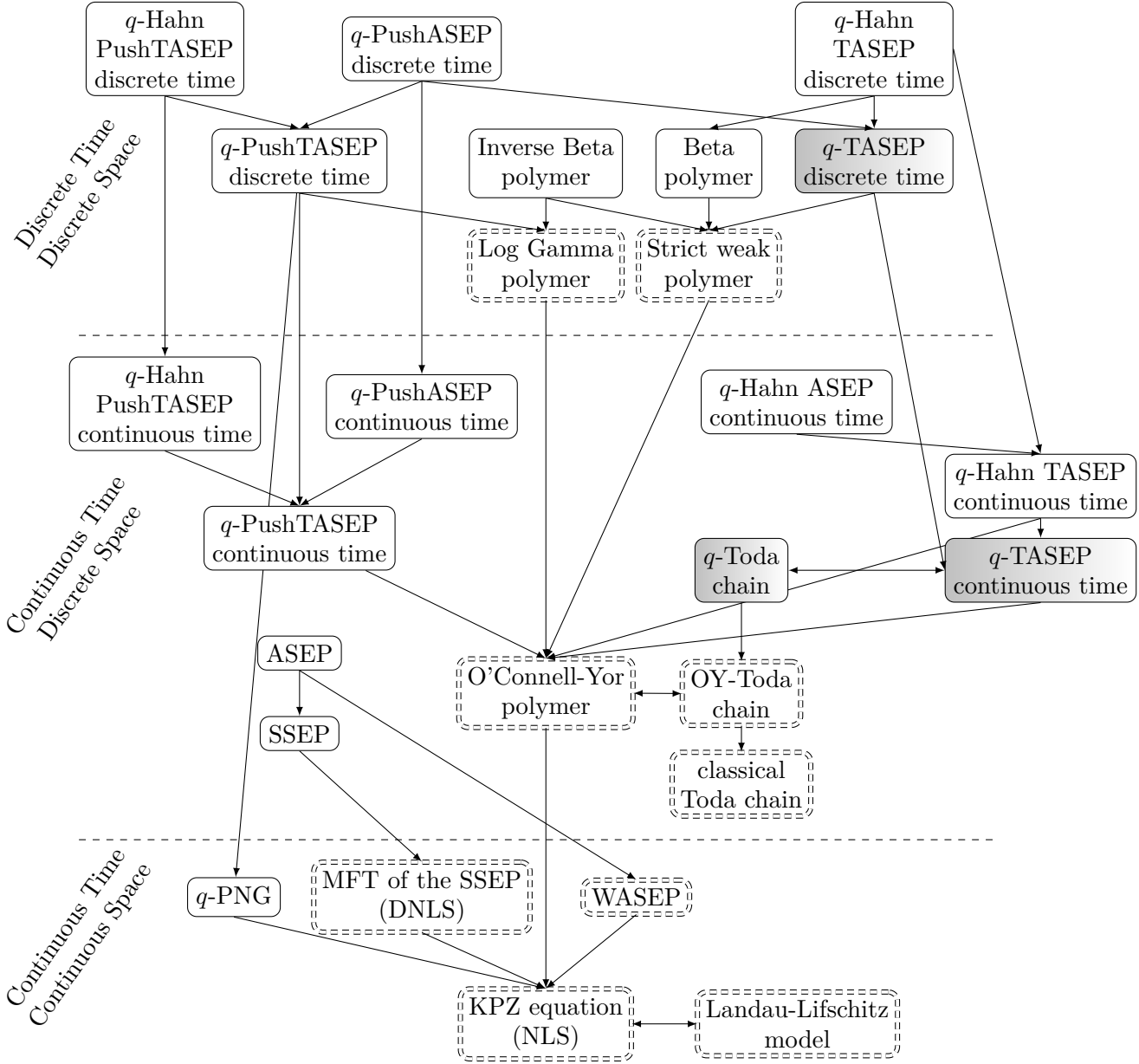

Although the present paper focused on the $q$-TASEP, additional models can be studied following our approach. As a guide for future work, we summarize in Figure~\ref{fig:phylogenetic-tree} the models within the KPZ class that are adjacent to the $q$-TASEP, and the models whose weak noise theory has been previously studied. The closest to study would be the $q$-Hahn models, see \cite{corwin2014macdonald,corwin-qhahn-tasep}, and the $q$-Push models, see \cite{CorwinqPushTasep}. As for the $q$-TASEP, we expect to unveil novel classical integrable models with this line of research.

Our results open multiple research directions. We have represented the saddle point non-linear systems in terms of $2\times 2$ Lax pairs. It is known, see e.g. \cite{Sklyanin1,Sklyanin2} and more recently \cite{krajenbrink2023weak} in the context of the O'Connell-Yor polymer, that there is a duality between $2\times 2$ Lax pairs and $N \times N$ Lax pairs if we study a particle system on a lattice of size $N$. Stating these dualities and deriving the large Lax pair is left for future work. As connexions exist between Lax matrices and diagonalisation algorithms, see \cite{deiftQRIntegrable}, we also expect that the novel Lax systems will be related to novel eigenvalue problems. Moreover, a Fredholm framework was developed in \cite{UsWNTDroplet2021,krajenbrink2023weak} 
to express the general solution of the weak noise equations using derivatives of Fredholm determinants and resolvent of scattering operators. Extending this framework to the $q$-TASEP remains to be done. 

Additionally, we have seen the appearance of classical dilogarithms 
(in the observable and in the rate functions)
in both the discrete and continuous time $q$-TASEP. We anticipate that dilogarithms will be ubiquitous for all weak noise theories involving $q$-deformed functions in the $q\to 1$ limit. It is interesting to note that they also arise in some large deviation 
problems related to random matrices 
\cite{fyodorov2024zeros}.
As the stochastic models in the KPZ class can also be represented by generalisations of the quantum delta Bose gas, it would be interesting to understand the role that dilogarithms play for the non-commuting quantum fields \cite{faddeev1994quantum}.

\newpage

\appendix

\section{Short Compendium on $q$-deformed functions}

\label{sec:compendium-q-math}

We recall the definition of the infinite $q$-Pochhammer symbol as
\begin{equation}
\label{eq:def-q-infinite-pochhammer}
    (a,q)_{\infty} =\prod_{\ell=0}^{\infty}(1-q^\ell a)
\end{equation}
as well as its finite $n$ version
\begin{equation}
\label{eq:def-q-finite-pochhammer}
    (a;q)_{n} = \frac{(a;q)_{\infty}}{(a q^n;q)_{\infty}}= \prod_{\ell=0}^{n-1}(1-aq^\ell)
\end{equation}
The $q$-factorial is defined as 
\begin{equation}
\label{eq:def-q-factorial}
[n]_q!=\frac{(q;q)_n}{(1-q)^n}
\end{equation}
The $q$-deformed binomial is defined as
\begin{equation}
\label{eq:def-q-binomial}
    \frac{(q;q)_m}{(q;q)_{m-j}(q;q)_j}=\binom{m}{j}_q
\end{equation}
See Ref.~\cite{corwin-qhahn-tasep} for more details about these formula.

We note another identity which could be used in the derivation of the first cumulant approximation. From Ref.~\cite{ito2014ramanujan}, we have the Ramanujan $_1 \psi_1$ summation formula: for $|q|<1, |b/a|<|z|<1$, $a\notin q^n, n\in\Z$, 
\begin{equation}
\sum_{n\in\Z} \frac{(bq^n;q)_\infty}{(aq^n;q)_\infty} z^n 
=
\frac{(az;q)_\infty (\frac{q}{az};q)_\infty (q;q)_\infty (\frac{b}{a};q)_\infty}{(a;q)_\infty (\frac{q}{a};q)_\infty (z;q)_\infty (\frac{b}{az};q)_\infty} . 
\label{ram}
\end{equation}

Specialising the Ramanujan's formula again to $a=-1/u, b=-q/u, z= v/\zeta$, 
\begin{equation}
\sum_{n\in\Z}\frac{1}{1+q^n/u} \left( \frac{v}{\zeta} \right)^n
=
\frac{(\frac{v}{-u \zeta})_\infty (\frac{-qu \zeta}{v};q)_\infty (q;q)_\infty^2}
       {(-1/u;q)_\infty (-qu;q)_\infty (v/\zeta;q)_\infty (q\zeta/v;q)_\infty}
\end{equation}

\section{Conserved quantities of the scattering theory of the continuous time $q$-TASEP}
\label{app:conserved}

In this Appendix we derive the conserved quantities and currents of the weak noise theory system of the
continuous time $q$-TASEP.  

\subsection{Ricatti equations}

Consider the equations \eqref{eq:scattering-linear-continuous-qtasep},
with the Lax pair given in \eqref{eq:lax-pair-continuous-q-tasep0}.
Let us first define the Ricatti variable $\Gamma$ and its inverse $\tilde{\Gamma}$ as
\begin{equation}
    \Gamma_n = \frac{v_n^{(2)}}{v_n^{(1)}}, \quad \tilde{\Gamma}_n=\frac{1}{\Gamma_n}
\end{equation}
where $v_n^{(1,2)}$ are the two components of the vector $v_n$. 
Dividing the two equations of the space part of the Lax pair, we obtain the following recursions for the Ricatti variable and its inverse

\begin{equation}
\label{eq:ricatti-eq-1}
   ( \Gamma_{n+1}(1-\lambda^2)+y_n)(z_n\Gamma_n-1)=(\Gamma_{n+1}-\Gamma_n) \lambda^2
\end{equation}
and
\begin{equation}
\label{eq:ricatti-eq-2}
   ( 1-\lambda^2+y_n\tilde{\Gamma}_{n+1})(z_n-\tilde{\Gamma}_{n})=(\tilde{\Gamma}_{n}-\tilde{\Gamma}_{n+1}) \lambda^2
\end{equation}
As we shall see subsequently, we need to expand these equations to obtain the Taylor and Laurent series of the Ricatti variables.

\subsection{Continuity equations}
The continuity equations will be obtained as the compatibility of the dynamics of $\log (v_n^{(1)}-v_n^{(2)})$ and $\log v_n^{(2)}$ respectively. We introduce the following notation for the finite difference $\Delta^+ f_n=f_{n+1}-f_n$. The compatibility will subsequently be obtained from the commutation relation 
\begin{equation}
    \p_\tau \Delta^+ = \Delta^+ \p_\tau
\end{equation}

\subsubsection{First continuity equation}
From the Lax pair system \eqref{eq:scattering-linear-continuous-qtasep}-\eqref{eq:lax-pair-continuous-q-tasep0}, we obtain the pair of equations for $\log v_n^{(2)}$


\begin{equation}
\label{eq:continuity-eq-rescaled-2}
    \begin{cases}
            \p_\tau \log (-v_n^{(2)}\lambda^{-n}e^{\frac{\lambda^2-1}{2}\tau})=-\frac{y_n z_{n-1}}{2}+y_{n}\tilde{\Gamma}_n\\
            \\
            \log\left(\frac{-v^{(2)}_{n+1}\lambda^{-n-1}e^{\frac{\lambda^2-1}{2}\tau}}{-v^{(2)}_{n}\lambda^{-n}e^{\frac{\lambda^2-1}{2}\tau}}\right) = \log(1+\frac{y_n z_n}{\lambda^2}-\frac{y_n}{\lambda^2}\tilde{\Gamma}_n ) -\frac{1}{2}\log (1+y_n z_n)
    \end{cases}
\end{equation}
which yields the following continuity equation
\begin{equation}
    \p_\tau  \left[ \log(1+\frac{y_n }{\lambda^2}(z_n-\tilde{\Gamma}_n) ) -\frac{1}{2} \log (1+y_n z_n)\right]= \Delta^+ \left[-\frac{y_n z_{n-1}}{2}+y_{n}\tilde{\Gamma}_n \right]
\end{equation}
We interpret $J_n^{(2)} =-\frac{y_n z_{n-1}}{2}+y_{n}\tilde{\Gamma}_n $ as a generalised current and $\varrho_n^{(2)}= \log(1+\frac{y_n z_n}{\lambda^2}-\frac{y_n}{\lambda^2}\tilde{\Gamma}_n ) -\frac{1}{2}\log (1+y_n z_n) $ as a generalised density. 

\begin{remark}
We first observe from an explicit computation that we have the conservation law
    \begin{equation}
    -\p_\tau \log(1+y_n z_n) = y_{n+1}z_n- y_n z_{n-1}=\Delta^+ (y_n z_{n-1})
\end{equation}
which could be used to further simplify the continuity equation.
\end{remark}
\subsubsection{Second continuity equation}
We now repeat the space part of the computation for $\log (v_n^{(1)}- v_{n}^{(2)})$. From the Lax pair equations \eqref{eq:scattering-linear-continuous-qtasep} \eqref{eq:lax-pair-continuous-q-tasep0}, we have
\begin{equation}
\label{eq:continuity-eq-rescaled-3}
    \log \left(\frac{e^{-\frac{\lambda^2-1}{2}\tau} \lambda^{n+1} (v_{n+1}^{(1)}- v_{n+1}^{(2)})}{e^{-\frac{\lambda^2-1}{2}\tau} \lambda^n (v_n^{(1)}- v_{n}^{(2)})} \right) = \log(1+y_n+((\lambda^2-1)z_n - \lambda^2-y_n z_n)\Gamma_n) - \log(1-\Gamma_n) - \frac{1}{2}\log (1+y_n z_n)
\end{equation}
The second equation will be used to obtain one of the scattering coefficient, see Eq.~\eqref{eq:scat-diff-jost}.

\subsubsection{Representation of the scattering coefficients $a(\lambda)$ and $\atilde(\lambda)$ with the Ricatti variables}

\begin{itemize}
    \item  With the particular choice of $\vec{v}_n = e^{\frac{\lambda^2-1}{2}\tau}\phi_n$, recalling the asymptotics \eqref{eq:def-scattering-amplitude} and \eqref{eq:scattering-coefficient-asymptotics}, we have (recalling $c=1)$


\begin{equation}
\label{eq:scat-diff-jost}
    a(\lambda)=\lim_{n\to \infty }e^{-\frac{\lambda^2-1}{2}\tau} \lambda^n (v_n^{(1)}-c v_{n}^{(2)})
\end{equation}
Since $\lim_{n\to -\infty} e^{-\frac{\lambda^2-1}{2}\tau} \lambda^n v_{n}^{(1)}=1$ and $\lim_{n\to -\infty} e^{-\frac{\lambda^2-1}{2}\tau} \lambda^n v_{n}^{(2)}=0$, we can sum \eqref{eq:continuity-eq-rescaled-3} over integers in $\Z$ and we obtain
\begin{equation}
\label{eq:conserved-q-scattering-a}
    \log a(\lambda)=\sum_{n=-\infty}^{+\infty} \left[ \log(1+y_n+((\lambda^2-1)z_n-\lambda^2-y_n z_n)\Gamma_n)-\log(1-\Gamma_n)-\frac{1}{2}\log (1+y_n z_n) \right] 
\end{equation}
and equivalently
\begin{equation}
    a(\lambda)=\prod_{n=-\infty}^{+\infty} \frac{1+y_n+((\lambda^2-1)z_n-\lambda^2-y_n z_n)\Gamma_n}{(1-\Gamma_n)\sqrt{1+y_nz_n}}
\end{equation}




\item  With the particular choice of $\vec{v}_n = e^{\frac{1-\lambda^2}{2}t}\bar{\phi}_n$ and summing the second equation of \eqref{eq:continuity-eq-rescaled-2} over integers in $\Z$ we obtain the expected relation between the scattering amplitude and this set of conserved charges as

\begin{equation}
\label{eq:conserved-q-scattering-atilde}
    \log \atilde(\lambda) = \sum_{n=-\infty}^{+\infty} \left[\log(1+\frac{y_n }{\lambda^2}(z_n-\tilde{\Gamma}_n) )-\frac{1}{2}\log (1+y_n z_n)\right]    \quad \longleftrightarrow \quad 
        \atilde(\lambda) = \prod_{n=-\infty}^{+\infty} \frac{1+\frac{y_n }{\lambda^2}(z_n-\tilde{\Gamma}_n) }{\sqrt{1+y_n z_n}}
    \end{equation}
\end{itemize}
\subsection{Conserved charges}
\label{subsec:supp-mat-conserved-quantities}
We now complete the determination of the conserved charges in the system by proceeding to the suitable expansion of the continuity equations. One can choose to expand either $\log a (\lambda)$ or $\log \tilde{a}(\lambda)$ around a particular choice of $\lambda$, consistent with the domain of analyticity of each function.

\subsubsection{Laurent expansion of $\atilde(\lambda)$}

Since $\log \atilde(\lambda)$ is analytic for large $\abs{\lambda}$, we choose to expand Eqs.~\eqref{eq:ricatti-eq-1}-\eqref{eq:ricatti-eq-2}-\eqref{eq:conserved-q-scattering-atilde} as  Laurent series
\begin{equation}
    \log \atilde(\lambda)=\sum_{\ell=0}^{\infty} \frac{\tilde{C}_\ell}{\lambda^{2\ell} }, \quad \Gamma_n = \sum_{\ell=0}^{\infty} \frac{\Gamma_n^{(\ell,\infty)}}{\lambda^{2\ell} }, \quad \tilde{\Gamma}_n = \sum_{\ell=0}^{\infty} \frac{\tilde{\Gamma}_n^{(\ell,\infty)}}{\lambda^{2\ell} }
\end{equation}
and
\begin{equation}
\varrho_n^{(2)}=\sum_{\ell=0}^{\infty} \frac{\varrho^{(2)}_{n,\ell}}{\lambda^{2\ell} }, \quad J_n^{(2)}=\sum_{\ell=0}^{\infty} \frac{J^{(2)}_{n,\ell}}{\lambda^{2\ell} }, \quad \tilde{C}_\ell = \sum_{n \in \Z}\varrho^{(2)}_{n,\ell}
\end{equation}


\begin{enumerate}
\item For $\ell=0$
\begin{equation}
    \label{eq:conserved-quantities-laurent-suppmat}
   \varrho^{(2)}_{n,0}=-\frac{1}{2} \log (1+y_n z_n), \quad J^{(2)}_{n,0}=\frac{y_n z_{n-1}}{2}
    \end{equation}
    \item For $\ell=1$
    \begin{equation}
   \varrho^{(2)}_{n,1}=y_n \left(z_n-z_{n-1}\right) , \quad J^{(2)}_{n,1}=y_n \left(z_{n-2}-z_{n-1}\right) \left(1+y_{n-1} z_{n-1}\right)
    \end{equation}
    \item For $\ell=2$
    \begin{equation}
   \varrho^{(2)}_{n,2}=\frac{1}{2} y_n \left(-y_n \left(z_{n-1}-z_n\right)^2-2 \left(z_{n-2}-z_{n-1}\right) \left(1+y_{n-1} z_{n-1}\right)\right)
   \end{equation}
   \begin{equation} J^{(2)}_{n,2}=y_n \left(-y_{n-2} z_{n-2}^2+y_{n-1} \left(z_{n-2}-z_{n-1}\right){}^2+z_{n-3} \left(y_{n-2} z_{n-2}+1\right)-z_{n-2}\right) \left(1+y_{n-1} z_{n-1}\right)
    \end{equation}
\end{enumerate}

The first conserved charges read
\begin{equation}
\begin{split}
\tilde{C}_0 &= -\frac{1}{2}\sum_{n =-\infty}^{\infty} \log(1+y_n z_n )\\
    \tilde{C}_1 &=\sum_{n =-\infty}^{\infty}y_n \left(z_n-z_{n-1}\right)    \\
    \tilde{C}_2  &= \sum_{n =-\infty}^{\infty}\frac{1}{2} y_n \left(-y_n \left(z_{n-1}-z_n\right)^2-2 \left(z_{n-2}-z_{n-1}\right) \left(1+y_{n-1} z_{n-1}\right)\right)
    \end{split}
\end{equation}
and one can verify using the dynamical equations for $\{ z_n, y_n \} \eqref{eq:sp-continuous-q-tasep-homogeneous}$  that for any solution we obtain  
\begin{equation}
 \p_\tau \varrho^{(2)}_{n,\ell}=J^{(2)}_{n+1,\ell}   - J^{(2)}_{n,\ell}
\end{equation}

\begin{remark}
    We further note that quantities $y_n \left(z_n-z_{n-1}\right) $ and $\left(y_{n+1}-y_n\right) z_n$ are a current and a density at the same time. 
\end{remark}



In the case of the step initial condition, the first conserved charge evaluated at initial and final times reads
    \begin{equation}
    \label{eq:ctilde0-conservedquantity}
        \tilde{C}_0 = -\frac{1}{2} \log (1+uz_N(\tau=T))= - \frac{1}{2}\sum_{n=1}^{+\infty}\log(1+y_n(\tau=0))
    \end{equation}

We  recall from \eqref{eq:continuous-qtasep-legendre-derivative} that we are interested in
\begin{equation}
\label{eq:legendreupsiplog}
    \Psi_N'(u)=\frac{\log(1+u z_N(\tau=T))}{u}
\end{equation}
We can thus relate the large deviation rate function $\Psi_N(u)$
to $\tilde{C}_0$ as

\begin{equation}
    \tilde{C}_0 
    =-\frac{1}{2}\log(1+u z_N(\tau=T))=-\frac{u\Psi_N'(u)}{2}
\end{equation}
Thus $\Psi_N(u)$ can be obtained from the knowledge of $\tilde a(\lambda)$ akin to the calculation introduced in \eqref{eq:riemann-hilbert-atilde-explicit}-\eqref{eq:riemann-hilbert-atilde-explicit-2}.

\begin{remark}
    One could also expand the Ricatti variables around $\lambda=1$ leading to other representations of the conserved charges.
\end{remark}

\section{More on the continuous time $q$-TASEP}
\label{app:more-qtasep}
\subsection{$q$-Toda chain as a gauge equivalent model}
It is possible to introduce a Toda counterpart to the $q$-TASEP model, similarly to what was done in Ref.~\cite{krajenbrink2023weak} between the weak noise theory of the Yor-O'Connell polymer and a generalisation of the Toda model. We thus introduce the same change of variable
\begin{equation}
\label{eq:change-q-tasep-toda-continuous}
  z_n(\tau)= e^{h_n(\tau)}, \qquad y_n(\tau)z_n(\tau)=\beta(p_n(\tau)+\gamma)
\end{equation}
which transforms in the homogeneous case the dynamics \eqref{eq:sp-continuous-q-tasep} onto
\begin{equation}
\label{eq:qtasep-toda-gauge-equiv}
    \begin{split}
        \p_\tau h_n&=(e^{h_{n-1}-h_n}-1) (1+\beta  (p_n+  \gamma) ) \\
        \p_\tau p_n&= \left(e^{h_{n-1}-h_n} (p_n+\gamma )-e^{ h_n-h_{n+1}} (p_{n+1}+\gamma )\right) (1+\beta 
(p_n+\gamma) )
    \end{split}
\end{equation}
Performing the change of variable \eqref{eq:change-q-tasep-toda-continuous} into the Lax pair of the continuous time $q$-TASEP \eqref{eq:lax-pair-continuous-q-tasep}, the Lax pair of the derived Toda model reads
\begin{equation}
\begin{split}
    &L_n=
 \begin{pmatrix}
            1 & 0 \\
            -\beta(p_n+\gamma)e^{-h_n}   & 1
        \end{pmatrix}
        \begin{pmatrix}
            \frac{1}{\lambda  \sqrt{\beta(p_n+\gamma) +1}} & 0 \\
            0 & \lambda  \sqrt{\beta(p_n+\gamma)+1}
        \end{pmatrix}
             \begin{pmatrix}
            1 & (\lambda^2-1)e^{h_n}    \\
            0 & 1
        \end{pmatrix}\\
        &= \begin{pmatrix}
            e^{\frac{h_n}{2}} & 0 \\
            0  & e^{-\frac{h_n}{2}}
        \end{pmatrix}\begin{pmatrix}
            1 & 0 \\
            -\beta(p_n+\gamma)   & 1
        \end{pmatrix}
        \begin{pmatrix}
            \frac{1}{\lambda  \sqrt{\beta(p_n+\gamma) +1}} & 0 \\
            0 & \lambda  \sqrt{\beta(p_n+\gamma)+1}
        \end{pmatrix}
             \begin{pmatrix}
            1 & (\lambda^2-1)   \\
            0 & 1
        \end{pmatrix}
        \begin{pmatrix}
            e^{-\frac{h_n}{2}} & 0 \\
            0  & e^{\frac{h_n}{2}}
        \end{pmatrix}\\
        U_{n}&=
    \begin{pmatrix}
 \frac{\lambda ^2-1}{2} +\frac{\beta(p_n+\gamma)e^{h_{n-1}-h_n} }{2} & (1-\lambda ^2) e^{h_{n-1}} \\
 \beta(p_n+\gamma)e^{-h_n} & -\frac{\lambda ^2-1}{2} -\frac{\beta(p_n+\gamma)e^{h_{n-1}-h_n} }{2} \\
\end{pmatrix}
    \end{split}
\end{equation}
The Lax matrices still verify $\Det L_n=1$ and $\Tr U_n=0$. This Toda-like system will converge to the one related to the WNT of the OY polymer, see Ref.~\cite[Section \textit{Limit to the classical Toda system}] {krajenbrink2023weak}. The limit from \eqref{eq:qtasep-toda-gauge-equiv} to the system obtained in \cite{krajenbrink2023weak} is obtained through the scaling \eqref{eq:scaling-limit-wnt-qtasep}.

\begin{remark}
    It is tempting to apply the gauge transformation \eqref{eq:gauge-semi-discrete} to this Lax pair with a diagonal gauge $g_n =\begin{pmatrix}
            e^{-\frac{h_{n-1}}{2}} & 0 \\
            0  & e^{\frac{h_{n-1}}{2}}
        \end{pmatrix}$ so that the system solely depends on the variables $(p_n, h_n-h_{n-1})$.
\end{remark}


Note that the Jacobian of the transformation \eqref{eq:change-q-tasep-toda-continuous} is constant which might be useful beyond the weak noise limit.

\subsection{Convergence of the continuous time $q$-TASEP to the O'Connell-Yor polymer}
As mentionned in the main text, the continuous time $q$-TASEP is known to converge towards the O'Connell-Yor polymer, see e.g., \cite{borodin2014duality,borodin2014macdonald}. We have provided
some details about this convergence in terms of (i) the stochastic
equation (ii) of our results for the large deviation functions of the two models  for some specific initial conditions (step to droplet).  
In this Appendix, for completeness we show the convergence (iii) at the level of
the dynamical field theories (iv) within the weak noise theory, directly from the non-linear system and from its Lax pair representation.

\subsubsection{Convergence of the field theory}
From Ref.~\cite[Eq.~(59)]{borodin2014duality}, the following scaling applied to the $q$-TASEP leads to its convergence to the Yor-O'Connell polymer model
\begin{equation}
\label{eq:scaling-qtasep-to-yo-2}
    q=e^{-\delta}, \quad t=\delta^{-2}s , \quad \mathsf{x}_n = \delta^{-2}\tau- (n-1)\delta^{-1}\log \delta^{-1}-\delta^{-1} \mathsf{F}_n(\tau), a_i = e^{-\delta \tilde{a}_i}
\end{equation}
From Eq.~\eqref{eq:MSR-continuous-q-tasep}, the $q$-TASEP is described by the following action
\begin{equation}
    S_0[z,\tilde{z}]=\sum_{n\geq 1} \int \rmd t \left[ \tilde{z}_n \p_t  z_n-a_n(z_n-z_{n-1}) \frac{e^{(q-1)z_n\tilde{z}_n}-1}{z_n} \right]
\end{equation}
Under the rescaling \eqref{eq:scaling-qtasep-to-yo-2}, the $q$-deformed partition function becomes (going to the field theory representation)

\be 
q^{{x}_n+n} ={ z}_n(t)\to \delta^{1-n} e^{- \frac{s}{\delta} + {F}_n(s) }\equiv\delta^{1-n} e^{- \frac{s}{\delta} + \frac{3 }{2}s  }{Z}_{n}^{\text{OY}}(s) \, ,
\ee 
its time derivative reads
\be 
\partial_t z_n = \delta^2 \partial_s ( \delta^{1-n} e^{- \frac{s}{\delta}+\frac{3}{2}s  } Z_{n}^{\text{OY}}(s) ) 
= \delta^{1-n} e^{- \frac{s}{\delta} +\frac{3}{2}s } ( (\frac{3 \delta^2}{2}- \delta)  Z_{n}^{\text{OY}}(s)  + \delta^2 \partial_s Z_{n}^{\text{OY}}(s) )  \, ,
\ee 
and the finite difference of the partition function reads
\be 
z_n - z_{n-1} = \delta^{1-n} e^{- \frac{s}{\delta}+\frac{3}{2}s } 
( Z_{n}^{\text{OY}}(s)
- \delta Z_{n-1}^{\text{OY}}(s) ) \, .
\ee 
We additionally rescale the response field as
\begin{equation}
    \tilde{z}_n(t) = \delta^{n-1}e^{s/\delta-\frac{3}{2}s}\tilde{Z}_n^{\text{OY}} (s)
\end{equation}
Overall the action gets transformed as
\begin{equation}
    \begin{split}
        &\sum_{n\geq 1} \int \rmd t \left[ \tilde{z}_n \p_t  z_n-a_n(z_n-z_{n-1}) \frac{e^{(q-1)z_n\tilde{z}_n}-1}{z_n} \right]\\
        \to & \sum_{n\geq 1} \int \rmd s \left[\tilde{Z}_n^{\text{OY}} \p_s Z_n^{\text{OY}}(s ) -(\tilde{a}(n)-1) Z_n^{\text{OY}}(s ) \tilde{Z}^{\text{OY}}_n(s )-\frac{1}{2} Z_n^{\text{OY}}(s )^2 \tilde{Z}^{\text{OY}}_n(s ){}^2-Z_{n-1}^{\text{OY}}(s ) \tilde{Z}^{\text{OY}}_n(s
   )\right]\\
    \end{split}
\end{equation}
which is identical to the action of the Yor-O'Connell polymer
\cite[Eq.~(S54)]{krajenbrink2023weak} 
and consistent with the stochastic equation \eqref{stocheqOY}.
Note that to obtain the proper action we had to expand the
exponential term in the $q$-TASEP action to quadratic order
(as Poisson variables become Gaussian in that limit).






\subsubsection{Convergence of the weak noise theory}
The weak noise theory of the O'Connell-Yor polymer was recently studied in Ref.~\cite{krajenbrink2023weak}. As we discussed in the text, its
large deviation rate function (for the point to point initial condition) 
can be recovered as a limit of the one derived
in this paper for the $q$-TASEP (for the step initial condition). 
This is in fact more general, and can
also be seen on the saddle point equations. 
Indeed starting from the weak noise theory of the space-inhomogeneous continuous time $q$-TASEP \eqref{eq:sp-continuous-q-tasep}
\be
\begin{split}
 \partial_t z_n &= a_n(z_{n-1}-z_n) (1 + y_n z_n) \\
  - \partial_t y_n &= (a_{n+1}y_{n+1}-a_n y_n) (1 + y_n z_n) 
 \end{split}
\ee
We define the rescaled variables as
\begin{equation}
\label{eq:scaling-limit-wnt-qtasep}
\begin{split}
    t&=\eta^{-1}\tilde{t}\\
    z_n(t)&=\eta^{1-n} e^{- \frac{\tilde{t}}{\eta}+\tilde{t} }Z_{n}^{\text{OY}}(\tilde{t})\\
     y_n(t) &= -\eta^{n}e^{\tilde{t}/\eta-\tilde{t}}\tilde{Z}_n^{\text{OY}}(\tilde{t})\\
     a_n &= e^{-\eta \tilde{a}_n}
\end{split}
\end{equation}
and readily obtain in the $\eta \to 0$ limit the space-inhomogeneous weak noise system of the OY polymer studied in \cite{krajenbrink2023weak}
\begin{equation}
\label{eq:YO-inhomogeneous-system}
    \begin{split}
        \p_{\tilde{t}} Z^{\text{OY}}_n&= Z^{\text{OY}}_{n-1}-Z^{\text{OY}}_n + \tilde{a}_nZ^{\text{OY}}_n+   (Z^{\text{OY}}_n)^2 \tilde{Z}^{\text{OY}}_n  \\
        -\p_{\tilde{t}} \tilde{Z}^{\text{OY}}_n&=\tilde{Z}^{\text{OY}}_{n+1}-\tilde{Z}^{\text{OY}}_{n}+ \tilde{a}_n\tilde{Z}^{\text{OY}}_n+   Z^{\text{OY}}_n (\tilde{Z}_n^{\text{OY}})^2 
    \end{split}
\end{equation}

\subsubsection{Convergence of the Lax pair}
In the space-homogeneous case, the Lax pair of the weak noise theory of the $q$-TASEP reads from \eqref{eq:lax-pair-continuous-q-tasep}
\begin{equation}
\begin{split}
L_n&
 =\begin{pmatrix}
            1 & 0 \\
            -y_n & 1
        \end{pmatrix}
        \begin{pmatrix}
            \frac{1}{\lambda  \sqrt{y_{n} z_n+1}} & 0 \\
            0 & \lambda  \sqrt{y_{n} z_n+1}
        \end{pmatrix}
             \begin{pmatrix}
            1 & (\lambda^2-1)z_n \\
            0 & 1
        \end{pmatrix}, \\ 
U_{n}&=
    \begin{pmatrix}
 \frac{\lambda ^2-1}{2} +\frac{y_{n} z_{n-1}}{2} & (1-\lambda ^2) z_{n-1} \\
 y_{n} & -\frac{\lambda ^2-1}{2} -\frac{y_{n} z_{n-1}}{2} \\
\end{pmatrix}
\end{split}
\end{equation}
The Lax pair of \eqref{eq:YO-inhomogeneous-system} was obtained in Ref.~\cite{krajenbrink2023weak}  as 
\begin{equation}
\label{eq:lax-pair-continuous-YO}
    \begin{split}
        L^{\text{OY}}_n &= \begin{pmatrix}
1  &   0 \\ 
-\tilde{Z}^{\text{OY}}_n &  1
\end{pmatrix}
\begin{pmatrix}
\frac{1}{\tilde{\lambda}} &   0 \\ 
0 &  \tilde{\lambda} - \frac{\tilde{a}_n}{\tilde{\lambda}}
\end{pmatrix}
\begin{pmatrix}
1  &   Z^{\text{OY}}_{n} \\ 
0 &  1
\end{pmatrix}, \quad U^{\text{OY}}_n=
\begin{pmatrix}
\frac{\tilde{\lambda}^2-1}{2}  &   -Z^{\text{OY}}_{n-1} \\ 
&\\
\tilde{Z}^{\text{OY}}_n &  \frac{1-\tilde{\lambda}^2}{2}
\end{pmatrix}
    \end{split}
\end{equation}
In order to study the convergence $(L_n,U_n)\to (L^{\text{OY}}_n,U^{\text{OY}}_n)$, one inserts the scaling limit \eqref{eq:scaling-limit-wnt-qtasep} of the $q$-TASEP inside the Lax pair, define the diagonal gauge matrix 
\begin{equation}
    g_n= \begin{pmatrix}
       \I  \eta^{-(n-1)/2}e^{-\frac{\tilde{t}}{2}(\frac{1}{\eta}-1)}& 0\\
        0& -\I \eta^{(n-1)/2}e^{\frac{\tilde{t}}{2}(\frac{1}{\eta}-1)}
    \end{pmatrix}
\end{equation}
and apply the gauge transformation \eqref{eq:gauge-semi-discrete}. Upon the rescaling of the spectral parameter $\lambda = \sqrt{\eta}\tilde{\lambda}$, one readily obtains the Lax pair \eqref{eq:lax-pair-continuous-YO} with $\tilde{a}_n=0$ in the limit $\eta \to 0$.

\begin{remark}
Note that the factor $\I$ in the definition of the gauge matrix has the sole role to change the sign of the off-diagonal elements of the Lax matrix of the OY polymer to keep the same conventions as in Ref.~\cite{krajenbrink2023weak}.    
\end{remark}


\section{Large deviation of polymer models through the first cumulant method}
\label{app:first-cumulant-method-polymers}

\subsection{Definition of the polymer models}

\subsubsection{Strict weak polymer}

The model was introduced in Ref.~\cite{corwin2015strict}.
The partition sums $Z_{n,t}$ of the strict weak polymer satisfy the recursion 
\begin{equation} \label{recursionstrict}
Z_{n,t+1} = Y_{n,t}  Z_{n,t} + Z_{n-1,t}
\end{equation}
where the random weights $Y_{n,t}$ are i.i.d. Gamma distributed random variables with distribution 
and characteristic function
\begin{equation}
\label{eq:gamma-distribution-properties}
Y \sim \mathrm{Gamma}(k,\theta), \quad P(Y) =\frac{Y^{k-1}}{\Gamma(k) \theta^k} e^{-\frac{Y}{\theta}} \quad , \quad \Var[Y]=k\theta^2   , \quad  \quad s >- \frac{1}{\theta}
\end{equation}

\subsubsection{Beta polymer}
The model was introduced in Ref.~\cite{barraquand2017random}.
The partition sum $Z_{n,t}$ of the point-to-point Beta polymer is defined by the recursion
\bea \label{recursionbeta} 
&& Z_{n,t} = Z_{n,t-1}  B_{n,t} + Z_{n-1,t-1}(1-B_{n,t}), \quad , \quad  \text{if } t \geq n \geq 1 \\
&& Z_{t+1,t} = Z_{t,t-1}, \quad , \quad \text{if } t > 0 \label{secondeq} 
\eea  
where the random weights are i.i.d Beta distributed, $B_{n,t} \sim \mathrm{Beta}(\alpha, \beta) $
with PDF and variance 
\begin{equation}
\label{eq:beta-distribution-properties}
P(B) = \frac{\Gamma(\alpha + \beta)}{\Gamma(\alpha)\Gamma(\beta)} B^{-1+\alpha} (1-B)^{-1+\beta} 
\quad , \quad \Var[B]=\frac{\alpha \beta}{(\alpha+\beta)^2(\alpha+\beta+1)}
\end{equation}
Note that the second equation \eqref{secondeq} is equivalent to setting $B_{t+1,t}=0$, 
and the last one is implied.



\begin{remark}[The Beta random walk]
This model is related to the Beta random walk defined as \cite{barraquand2017random}, see also \cite{krajenbrinkhartmannbeta}
 \be
 P(x,t+1)=P(x-1,t) B_{x-1,t} + P(x+1,t) (1- B_{x+1,t}) \quad , \quad P(x,0)=\delta_{x0} 
 \ee
 where $P(x,t)$ is the probability that the walker is on site $x \in \mathbb{Z}$ at time 
 $t$. It is related to the point-to-point Beta polymer by
\be 
 Z_{n,t}= P(t-2 n + 2,t) \quad , \quad t+1 \geq n
 \ee 
\end{remark}

\subsubsection{Log Gamma polymer}

The model was introduced in 
Ref.~\cite{seppalainen2012scaling}.
The partition sums $Z_{n,m}$ of the Log Gamma polymer satisfy the recursion 
\be \label{recursionloggamma}
Z_{n,m} = w_{n,m} ( Z_{n-1,m}  + Z_{n,m-1} ) 
\ee 
where the random weights $w_{n,m}$ are i.i.d. inverse Gamma distributed random variables with PDF
and variance
\be 
\label{eq:inversegamma-distribution-properties}
w_{ij} = \mathrm{Inverse Gamma}(\gamma_{i,j}=\alpha_i + \beta_j), \quad P(w)=\frac{w^{-1-\gamma} }{\Gamma(\gamma)} e^{-\frac{1}{w}} ,\quad \Var[w]=\frac{1}{(\gamma-1)(\gamma-2)^2}
\ee 

\subsection{Fredholm determinant formula for the point-to-point polymer partition functions}

We now recall the Fredholm determinant formula which have been proved for
the case of the point to point polymers. We first recall the corresponding initial
conditions, as well as the observable which appears in the formula given below,
for each of the three models.

\subsubsection{Initial condition and observable for the point to point polymers}

\begin{itemize}
    \item Strict weak polymer. The point to point polymer is defined \cite{corwin2015strict} by
    the recursion \eqref{recursionstrict} with
    the initial condition $Z_{n,0} = \delta_{n,1}$ which implies the discrete light cone $Z_{n \leq 0,t}=0$ and $Z_{n \geq  t+1,t}=0$. One computes $Z_{n,t}$
    for $t \geq 0$ and the observable is $Z_{N,T}$ with $ 1 \leq N \leq T$.\\
    
    \item Beta polymer. The point to point polymer is defined  
    \cite{barraquand2017random} the recursion \eqref{recursionbeta}
    for $t \geq 0$
with initial condition $Z_{n,0} = \delta_{n1} $. It implies the discrete light cone $Z_{n \leq 0,t} = 0$, 
and $Z_{n,t} = 0$ for $n>t+1$, as well as $Z_{1,t} = Z_{1,t-1} B_{1,t}$
for $t>0$. The observable is $Z_{N,T}$ with $N \leq T+1$. \\

 \item Log Gamma polymer. The point-to-point polymer is defined \cite{borodin2013log} by 
    the recursion \eqref{recursionloggamma}  defined on the quadrant for $n\geq 1$ and $m\geq 1$ together with the boundary conditions $Z_{n,0} = \delta_{n,1}$ and $Z_{0,m}=0$
(one may consider that $Z_{n,m\leq 0} = 0$ and $Z_{n \leq 0,m}=0$). The observable is $Z_{N,M}$ with $N \geq M$. 
\end{itemize}


\subsubsection{Fredholm determinant formula}

The main theorem which encompasses the three cases takes the form

\begin{theorem}
    Consider $u \in \C$ such that $\Re u>0$ and a random variable $\mathtt{Z}$ defined below for each polymer model.
    Then one has 
    \be 
\label{eq:fred-det-strict-weak}
\mathbb{E} \big[ e^{- u \mathtt{Z}} \big] = {\rm det}(I + K_u)_{L^2(\mathtt{C})}
\ee 
where the kernel is
\be 
K_u(v,v') = \int_{\mathtt{D}}\frac{\rmd z}{2 \I \pi} \frac{\pi}{\sin(\pi (v-z))}  u^{z-v}
\frac{g(v)}{g(z)} \frac{1}{z -v'} 
\label{kernel-polymers} 
\ee 
\end{theorem}
In each case the random variable and the contours are
\begin{itemize}
    \item Strict weak polymer \cite[Theorem~1.7]{corwin2015strict}. 
    We have $\mathtt{Z}=Z_{N,T}$ for $N \leq T$, the weight $g(v)$ reads
    \be
\label{eq:g-func-strict-weak}
g(v) = g^{\rm SW}(v) = \frac{ \Gamma(v)^N}{\Gamma(k+v)^T} \theta^{-(T - (N-1)) v}
\ee 
and the contours are $\mathtt{C}=C_0$, a small positively oriented circle containing $0$ and 
$\mathtt{D}= \frac{1}{2}+ \I \R$.\\

\item Beta polymer \cite[Theorem 1.12]{barraquand2017random}. We have  $\mathtt{Z}=Z_{N,T}$ for $N,T \geq 0$, $N \leq T+1$, $\alpha>0$, $\alpha+\beta>0$, 
(note that for the Beta polymer $\mu=\alpha$ and $\nu=\alpha+\beta$ are also used in \cite{barraquand2017random}) and  the weight $g(v)$ reads
\be
\label{eq:g-func-beta-polymer}
g(v) = g^{\rm BP}(v) = \left(\frac{\Gamma(v)}{\Gamma(\alpha + \beta + v)}\right)^N \left(\frac{\Gamma(\alpha+\beta + v)}{\Gamma(\alpha  + v)}\right)^T \Gamma(\alpha + \beta + v) 
\ee 
and the contour $\mathtt{C}=C_{0}$ is a positively oriented closed contour enclosing $0$ but not $-\alpha - \beta$ nor $-1$ and $\mathtt{D}= \frac{1}{2}+ \I \R + v$ (the kernel is usually
written in terms of the integration variable $s=z-v$).

\item Log Gamma polymer \cite[Corollary 1.8]{borodin2013log}  -- \cite[Supp. Mat Section II.2]{barraquand2020stochastic}. We have $\mathtt{Z}=Z_{N,M}$ for $N \geq M$ 

\be
\label{eq:g-func-loggamma-polymer}
g(v) = g^{\rm LG}(v) = \frac{ \prod_{j=1}^M \Gamma(\beta_j+v) }{\prod_{j=1}^N \Gamma(\alpha_i - v)}
\ee 
The contour $\mathtt{C}=C_{\delta}$ is a positively oriented closed contour enclosing the set of $- \beta_j$ and no other singularity, the integration contour being $\mathtt{D}= \delta + \I \R$ such that $\delta < \alpha_j$, the contour $C_{\delta}$ lies to the left of 
$\delta + \I \mathbb{R}$ and all the poles at $z = v + 1,v + 2,...$ lie to the right of $\delta + \I \mathbb{R}$.
\end{itemize}





\begin{remark}
From the Beta polymer, 
one obtains \cite[Theorem 1.13]{barraquand2017random} the same result for the Beta random walk 
with $\mathtt{ Z}= P(X,T)$ 
and
\be 
g(v) = g^{\rm BRW}(v) = \left(\frac{\Gamma(v)}{\Gamma(\alpha +  v)}\right)^{\frac{T-X}{2}}  \left(\frac{\Gamma(\alpha+\beta + v)}{\Gamma(\alpha  + v)}\right)^{\frac{T+X}{2}}  
\Gamma(v) 
\ee  
which is identically to $g^{\rm BP}(v)$ using $X=T-2 N + 2$. 
\end{remark}

\subsection{Weak noise theory of the polymer models}

\subsubsection{Weak noise scaling}

We will consider the above polymer models in the weak noise limit
where the variance of the noise is small. Note however that the
space and time are not rescaled

\begin{itemize}
\item For the strict weak polymer one considers
\be 
\label{eq:rescaling-wnt-sw}
k=1/\varepsilon, \; \theta= \tilde \theta \varepsilon, \; \varepsilon \ll 1
\ee 
where $\tilde \theta$ remains fixed.\\

\item For the Beta polymer one considers 
\begin{equation}
\label{eq:rescaling-wnt-beta}
    \alpha=\tilde \alpha/\varepsilon, \, \beta=\tilde \beta/\varepsilon, \, \varepsilon \ll 1
\end{equation}
where $\tilde \beta$ remains fixed. The same scaling holds for the Beta random walk.
Note this scaling is different from the one considered in \cite{krajenbrinkhartmannbeta} where
space and time where rescaled (while here $X,T$ are fixed). \\
\item For the Log gamma polymer one considers
\begin{equation}
\label{eq:rescaling-wnt-lg}
     \alpha_i=\tilde \alpha_i/\varepsilon, \; \beta_j=\tilde \beta_j/\varepsilon , \, \varepsilon \ll 1
\end{equation}
where $\tilde \alpha$ and $\tilde \beta$ remain fixed.
\end{itemize}

\subsubsection{Large deviation functions} 

We now take the weak noise limit of the above Fredholm formula.
This leads to a large deviation form for the Laplace transform
of the PDF of the point to point partition functions.
We first give the result for each model 
and sketch the derivation below.

\begin{itemize}
\item For the strict weak polymer we obtain for $N \leq T$, for fixed $\Lambda$ 
\be 
\mathbb{E} \left[ e^{-  \frac{\Lambda }{\varepsilon} Z_{N,T} } \right] \underset{\varepsilon \ll 1}{\sim}  e^{-  \frac{1}{\varepsilon} \Psi^{\rm SW}(\Lambda) } 
\ee 
with
\begin{equation} \label{resultPsiSW} 
   \Psi^{\rm SW}(\Lambda) = - \int_{C_0'} \frac{\rmd v}{2\I \pi} \, \mathrm{Li}_2\left(-\Lambda v^{-N} (1+v)^{T}  \tilde \theta^{1-N+T} \right) \, .
\end{equation}
where $C'_0$ is a positively oriented circle containing 0.
\item For the Beta polymer, we obtain for $N \leq T+1$,
\be 
\mathbb{E} \left[ e^{-  \frac{\Lambda }{\varepsilon} Z_{N,T} } \right] \underset{\varepsilon \ll 1 }{\sim}  e^{- \frac{1}{\varepsilon} \Psi^{\rm BP}(\Lambda)} 
\ee 
with
\begin{equation} \label{resultPsiBP} 
   \Psi^{\rm BP}(\Lambda) = - \int_{C_0'} \frac{\rmd v}{2\I \pi} \, \mathrm{Li}_2\left(-\Lambda \left(\frac{v}{\tilde{\alpha}+\tilde{\beta}+v}\right)^{-N} \left(\frac{\tilde{\alpha}+v}{\tilde{\alpha}+\tilde{\beta}+v} \right)^{T}  (\tilde{\alpha}+\tilde{\beta}+v)^{-1} \right) \, .
\end{equation}
where $C'_0$ is a positively oriented closed contour enclosing $0$ but not $-\tilde{\alpha} - \tilde{\beta}$.\\

For the Beta random walk, the result extends replacing $Z_{N,T}$ by $P(X,T)$ and performing the change of variable $X=T-2 N + 2$.
\item For the Log Gamma polymer, for $N \geq M$ we rescale the partition function as
\begin{equation}
    Z_{N,M}=\varepsilon^{N+M-1}z_{N,M}
\end{equation}
so that
\be 
\mathbb{E} \left[ e^{-   \frac{\Lambda}{\varepsilon }  z_{N,M} } \right] \underset{\varepsilon \ll 1 }{\sim}  e^{-  \frac{1}{\varepsilon} \Psi^{\rm SW}(\Lambda) } 
\ee 
with
\begin{equation}
   \Psi^{\rm LG}(\Lambda) = - \int_{C_0'} \frac{\rmd v}{2\I \pi} \, \mathrm{Li}_2\left(-\Lambda \frac{(\tilde \alpha - v)^{N} }{(\tilde \beta+v )^{M}}  \right) \, .
\end{equation}
The generalisation to inhomogeneities is given by
\begin{equation} \label{resultPsiLG}
   \Psi^{\rm LG}(\Lambda) = - \int_{C_0'} \frac{\rmd v}{2\I \pi} \, \mathrm{Li}_2 \left(-\Lambda \frac{\prod_{i=1}^N(\tilde \alpha_i - v)}{ \prod_{j=1}^M(\tilde \beta_j+ v)  }\right) 
\end{equation}
where $C'_0$ is a positively oriented closed contour enclosing the set of $-\tilde{\beta}_j$.
\end{itemize}

These results are obtained by the first cumulant method. Since the kernel $K_u$ 
has the same form as in the case of the OY polymer, the 
steps are very similar and we only sketch the derivation, see 
\cite[Section~XVIII]{krajenbrink2023weak} for details. 
The only difference being the explicit form of the
function $g(v)$ we give its asymptotics form below for
each model.\\

The first cumulant method applied to kernels of the form \eqref{kernel-polymers}
gives generically the following asymptotic formula for the Fredholm determinant
\be 
{\rm det}(I + K_u)_{L^2(\mathtt{C})} \underset{\varepsilon \ll 1}{\sim} e^{-\frac{1}{\varepsilon} \Psi(\Lambda)} 
\ee 
where 
\begin{equation}
   \Psi(\Lambda) = - \int_{C_0'} \frac{\rmd v}{2\I \pi} \, \mathrm{Li}_2(-\Lambda e^{-\phi'(v)}) \, .
\end{equation}
where the argument of the dilogarithm is obtained from the limit
\begin{equation}
    \phi(v)=\lim_{\varepsilon \to 0}\varepsilon \log g\left(\frac{v}{\varepsilon}\right)
\end{equation}
which implicitly includes the rescaling of the model parameters (in the present cases \eqref{eq:rescaling-wnt-sw}, \eqref{eq:rescaling-wnt-beta} and \eqref{eq:rescaling-wnt-lg}) as well as
a rescaling of the Laplace parameter $u$ as
\begin{equation}  \label{rescalingu} 
    u= \frac{\Lambda}{\varepsilon^{1+\sigma}},  \quad\mathtt{Z}=\varepsilon^{\sigma}\mathtt{z}
\end{equation}
so that Eq. \eqref{eq:fred-det-strict-weak} becomes in the limit
\begin{equation}
    \mathbb{E}\left[e^{-\frac{\Lambda}{\varepsilon} \mathtt{z}} \right] \underset{\varepsilon \ll 1 }{\sim}  e^{-\frac{1}{\varepsilon}\Psi(\Lambda)}
\end{equation}

Let us detail now the calculation of the function $\phi(v)$ for
each model, as well as the rescaling factor $\varepsilon^\sigma$.

\begin{itemize}
    \item For the strict-weak polymer the function $g(v)$ is given by Eq.~\eqref{eq:g-func-strict-weak}
    and one finds     
\be
\begin{split}
& \varepsilon \log g^{\rm SW}_{ 1/\varepsilon,\tilde \theta \varepsilon}(\frac{v}{\varepsilon}) 
= N \left( v \log v + v (\log \tilde \theta -1 ) \right) - T \left( (1+v) \log(1+v) - (1+v) + v \log \tilde \theta \right) \\ \label{secondline} 
& - v \log \tilde \theta + ( T - v ) \log \varepsilon + \mathcal{O}(\varepsilon, \varepsilon \log \varepsilon) 
\end{split}
\ee
Taking a derivative one obtains 
\be 
\phi'(v) = N \log v - T \log(1+v) + (N-T-1) \log \tilde \theta 
\ee 
which leads to \eqref{resultPsiSW}. Note that we 
have discarded the term $- v \log \varepsilon$ in \eqref{secondline} which is absorbed in
the rescaling \eqref{rescalingu} of the Laplace parameter (with $\sigma=0$).
Indeed it is the term $u^{-v} g(v)$ which appears in the kernel. \\

\item 
For the Beta polymer, the function $g(v)$ is given by Eq.~\eqref{eq:g-func-beta-polymer} and one finds 
\bea  
 &&   \varepsilon \log g^{\rm BP}_{1/\varepsilon,\tilde{\beta}/\varepsilon}(\frac{v}{\varepsilon})=
    N (v \log v - v ) + (T-N+1) ((\tilde{\alpha}+\tilde \beta +v) \log(\tilde{\alpha}+\tilde \beta +v)-(\tilde{\alpha} + \tilde \beta +v)) \nn \\
 &&   - T ((\tilde{\alpha}+v) \log(\tilde{\alpha}+v)-(\tilde{\alpha}+v)) - v \log \varepsilon + (\tilde{\alpha}(N-1)+\tilde{\beta}(N-T-1)) \log \epsilon+\mathcal{O}(\varepsilon \log \varepsilon) \nonumber \\
 && 
\eea 
Hence, discarding again the term $- v \log \varepsilon$ which is absorbed
in the rescaling of the Laplace parameter (with $\sigma=0)$, one finds 
\be 
\phi'(v) = N \log v + (T-N+1)  \log(\tilde \alpha + \tilde \beta +v) - T \log(\tilde{\alpha}+v)
\ee 
which leads to \eqref{resultPsiSW}.



\item For the Log Gamma polymer, the function $g(v)$ is given by Eq.~\eqref{eq:g-func-loggamma-polymer} 
One finds, with $\alpha_i=\tilde \alpha_i/\varepsilon$ and $\beta_j=\tilde \beta_j/\varepsilon$ that
\begin{equation}
\begin{split}
  \varepsilon \log g^{\rm LG}_{\tilde{\alpha}/\varepsilon,\tilde{\beta}/\varepsilon}(\frac{v}{\varepsilon})&=\sum_{j=1}^M (\tilde{\beta}_j +v) (\log
   (\tilde{\beta}_j +v)-1)+\sum_{i=1}^N (v-\tilde{\alpha}_i ) (\log (\tilde{\alpha}_i -v)-1)\\
  &- (\sum_{j=1}^M (\tilde{\beta_j} +v)+\sum_{i=1}^N (v-\tilde{\alpha}_i ))(\log (\varepsilon ))
   \end{split}
\end{equation}
Hence, discarding the term $-(N+M) v \log \varepsilon$ which is absorbed
in the rescaling of the Laplace parameter (with now $\sigma=N+M-1)$, one finds 
\be 
\phi'(v) =  \sum_{j=1}^M\log(v + \tilde \beta_j) - \sum_{i=1}^N \log(\tilde \alpha_i - v) 
\ee 
which recovers \eqref{resultPsiLG}.



\end{itemize}

\bibliographystyle{unsrt}
\bibliography{discrete-wnt-modified-4.bib}

\end{document}

%% file: tikzgraphs.tex
\newcommand{\qTASEPContinuousLattice}{
\begin{tikzpicture}[
    node distance=0.75cm,
    particle/.style={circle, draw=black, fill=white, thick, minimum size=2mm, inner sep=0pt},
    particle_black/.style={circle, draw=black, fill=black, thick, minimum size=2mm, inner sep=0pt},
    arrow_line/.style={->, >=Stealth, thick, bend left},
    both_arrow_line/.style={->, >=Stealth, thick}
]

    \foreach \x in {0,...,6} {
        \ifthenelse{\x = 2 \OR \x = 5}{
            \node[particle_black] (P\x) at (\x*0.75,0) {}; 
        }{
            \node[particle] (P\x) at (\x*0.75,0) {}; 
        }
    }

    \draw[both_arrow_line] ([xshift=-1cm]P0.west) -- ([xshift=1cm]P6.east);

    \node[below=0.1cm of P2] {$x_i(t)$};

    \node[below=0.1cm of P5] {$x_{i-1}(t)$};

    \draw[arrow_line] (P2.north) to (P3.north) node[above=2mm] {rate $a_i(1 - q^{\text{gap}_i})$};

\end{tikzpicture}
}

\newcommand{\qTASEPDiscreteLattice}{\begin{tikzpicture}[
    node distance=0.75cm,
    particle/.style={circle, draw=black, fill=white, thick, minimum size=2mm, inner sep=0pt},
    particle_black/.style={circle, draw=black, fill=black, thick, minimum size=2mm, inner sep=0pt},
    arrow_line/.style={->, >=Stealth, thick, bend left=60},
    both_arrow_line/.style={->, >=Stealth, thick}
]

    \foreach \x in {0,...,6} {
        \ifthenelse{\x = 0 \OR \x = 4 \OR \x = 6}{
            \node[particle_black] (G\x) at (\x*0.75,0) {}; 
        }{
            \node[particle] (G\x) at (\x*0.75,0) {}; 
        }
    }

    \draw[both_arrow_line] ([xshift=-1cm]G0.west) -- ([xshift=1cm]G6.east);

    \node[below=0.1cm of G0] {$x_i(t)$};
    \node[below=0.1cm of G4] {$x_{i-1}(t)$};

    \draw[arrow_line] (G0) to (G1);
    \draw[arrow_line] (G0) to (G2);
    \draw[arrow_line] (G0) to (G3);
    \draw[arrow_line] (G4) to (G5);

    \node[above=5mm of G0.north] {$\mathbf{p}_{\text{gap}_i, a_i \alpha_{t+1}}(j)$};
    \node[above=5mm of G5.north] {$\mathbf{p}_{\text{gap}_{i-1}, a_{i-1} \alpha_{t+1}}(j)$};

\end{tikzpicture}}

\newcommand{\IntegrabilityCurvature}{
\begin{tikzpicture}[scale=3.5]
\node[below,font=\Large] at (0.5,0) {$L_{n,t}$};
\node[left,font=\Large] at (0,0.5) {$U_{n,t}$};
\node[right,font=\Large] at (1,0.5) {$U_{n+1,t}$};
\node[above,font=\Large] at (0.5,1) {$L_{n,t+1}$};

\node[below left] at (0,0) {$(n,t)$};
\node[above left] at (0,1) {$(n,t+1)$};
\node[below right] at (1,0) {$(n+1,t)$};
\node[above right] at (1,1) {$(n+1,t+1)$};

\draw[->,>=latex',decorate, decoration={snake, amplitude=.6mm, segment length=5mm}] (0,0) -- (1,1) node[midway, sloped, above] {compatibility};

\draw[->, dashed, line width=0.4mm,>=latex'] (0,0) -- (0,1);
\draw[->, dashed, line width=0.4mm,>=latex'] (0,1) -- (1,1);
\draw[->, line width=0.4mm,>=latex'] (0,0) -- (1,0);
\draw[->, line width=0.4mm,>=latex'] (1,0) -- (1,1);

\end{tikzpicture}
}